\documentclass[aps,twocolumn,nofootinbib,showpacs,superscriptaddress,prl]{revtex4-1}

\usepackage{hyperref}
\usepackage{ulem}
\usepackage{graphics}
\usepackage{epstopdf}
\usepackage{bm}
\usepackage{bbm}
\usepackage{amssymb}
\usepackage{epstopdf}
\usepackage{graphicx}
\usepackage[caption=false]{subfig}
\usepackage[cmex10]{amsmath}
\usepackage{amsthm}
\usepackage{color}
\usepackage{array}
\newcolumntype{C}[1]{>{\centering\let\newline\\\arraybackslash\hspace{0pt}}m{#1}}

\newcommand{\ket}[1]{\left|#1\right\rangle}
\newcommand{\kets}[2]{\left|#1\right\rangle_{_{#2}}}
\newcommand{\keti}[1]{|#1\rangle}
\newcommand{\bra}[1]{\left\langle#1\right|}

\begin{document}

\title{Simultaneous entanglement swapping\\of multiple orbital angular momentum states of light}

\author{Yingwen \surname{Zhang}$^{*}$}
\affiliation{CSIR National Laser Centre, PO Box 395, Pretoria 0001, South Africa}

\author{Megan \surname{Agnew}$^{*}$}
\affiliation{IPaQS, SUPA, Heriot-Watt, Edinburgh EH14 4AS, United Kingdom}
\thanks{These authors contributed equally to this work.}

\author{Thomas \surname{Roger}}
\affiliation{IPaQS, SUPA, Heriot-Watt, Edinburgh EH14 4AS, United Kingdom}

\author{Filippus S. \surname{Roux}}
\affiliation{CSIR National Laser Centre, PO Box 395, Pretoria 0001, South Africa}
\affiliation{School of Physics, University of Witwatersrand, Johannesburg 2000, South Africa}

\author{Thomas \surname{Konrad}}
\affiliation{School of Chemistry and Physics, University of KwaZulu-Natal, Private Bag X54001, Durban 4000, South Africa}
\affiliation{National Institute of Theoretical Physics, University of KwaZulu-Natal, Private Bag X54001, Durban 4000, South Africa}

\author{Daniele \surname{Faccio}}
\affiliation{IPaQS, SUPA, Heriot-Watt, Edinburgh EH14 4AS, United Kingdom}

\author{Jonathan \surname{Leach}}
\affiliation{IPaQS, SUPA, Heriot-Watt, Edinburgh EH14 4AS, United Kingdom}

\author{Andrew \surname{Forbes}}
\affiliation{CSIR National Laser Centre, PO Box 395, Pretoria 0001, South Africa}
\affiliation{School of Physics, University of Witwatersrand, Johannesburg 2000, South Africa}

\maketitle

\textbf{Entanglement swapping generates remote quantum correlations between particles that have not interacted and is the cornerstone of long-distance quantum communication, quantum networks, and fundamental tests of quantum science. In the context of spatial modes of light, high-dimensional entanglement provides an avenue to increase the bandwidth of quantum communications and provides more stringent limits for tests of quantum foundations. Here we simultaneously swap the entanglement of multiple orbital angular momentum states of light. The system is based on a degenerate filter that cannot distinguish between different anti-symmetric states, and thus entanglement swapping occurs for several thousand pairs of spatial light modes simultaneously.
}

An integral part of a quantum repeater is the ability to entangle two systems that have not interacted -- a process referred to as entanglement swapping \cite{Briegel1998,Pan1998,Shi2000,Jennewein2001,deRiedmatten2005,Kaltenbaek2009,Ma2012}. In optics, it is accomplished by interfering two photons via Hong-Ou-Mandel (HOM) interference \cite{Peeters2007,Nagali2009,DiLorenzoPires2010,Zhang2016}, each from a different entangled pair, in such a way that  their remote partners become mutually entangled. This allows the establishment of entanglement between two distant points without requiring single photons to travel the entire distance, thus reducing the effects of decay and loss.

While quantum communication has largely been demonstrated using two-level systems -- qubits -- to carry information, the use of high-dimensional systems allows more information to be encoded per particle. One way to accomplish this is to encode the information in the orbital angular momentum (OAM) of a photon. It is routinely possible to obtain OAM states entangled in very high dimensions \cite{Leach2010,Agnew2011,Dada2011,Krenn2014,Bolduc2016}, and entanglement of OAM is easily produced via spontaneous parametric downconversion (SPDC) \cite{Mair2001,Walborn2004}, making OAM an ideal method to increase information capacity \cite{Mafu2013,Mirhosseini2015}. Other high-dimensional systems that could increase information capacity include time bins \cite{Marcikic2002}, the path degree of freedom in waveguides \cite{Matthews2009}, and hybrid entanglement \cite{Barreiro2008,Karimi2010,Wang2015,Graham2015}. Recently, a number of multi-photon OAM experiments have been reported, including a demonstration of four-photon entanglement \cite{Hiesmayr2016} and the creation of Greenberger-Horne-Zeilinger states \cite{Malik2016}. However, realising entanglement swapping and teleportation in high dimensions has been thought to require increasing the photon number with dimension \cite{Goyal2013,Goyal2014}, a prohibitive constraint due to the low count rates associated with many-photon entanglement experiments.

In this work, we perform the first implementation of entanglement swapping of spatial states of light. We use photons entangled in the OAM degree of freedom and transfer entanglement from one pair of entangled photons to another, even though the final entangled pair have not interacted with each other. We present results for swapped entanglement in six two-dimensional subspaces. Four of these subspaces did not show entanglement prior to the entanglement swapping. We combine these six subspaces into a four-dimensional mixed state that is representative of the final state in high dimensions. We outline entanglement purification schemes to convert this mixed state into a pure high-dimensional state, allowing scalability of our approach to any dimension without the need for additional ancillary photons, thus providing a new approach towards high-dimensional, long-distance secure quantum communication.

\begin{figure*}
\centering
\includegraphics[width=\linewidth]{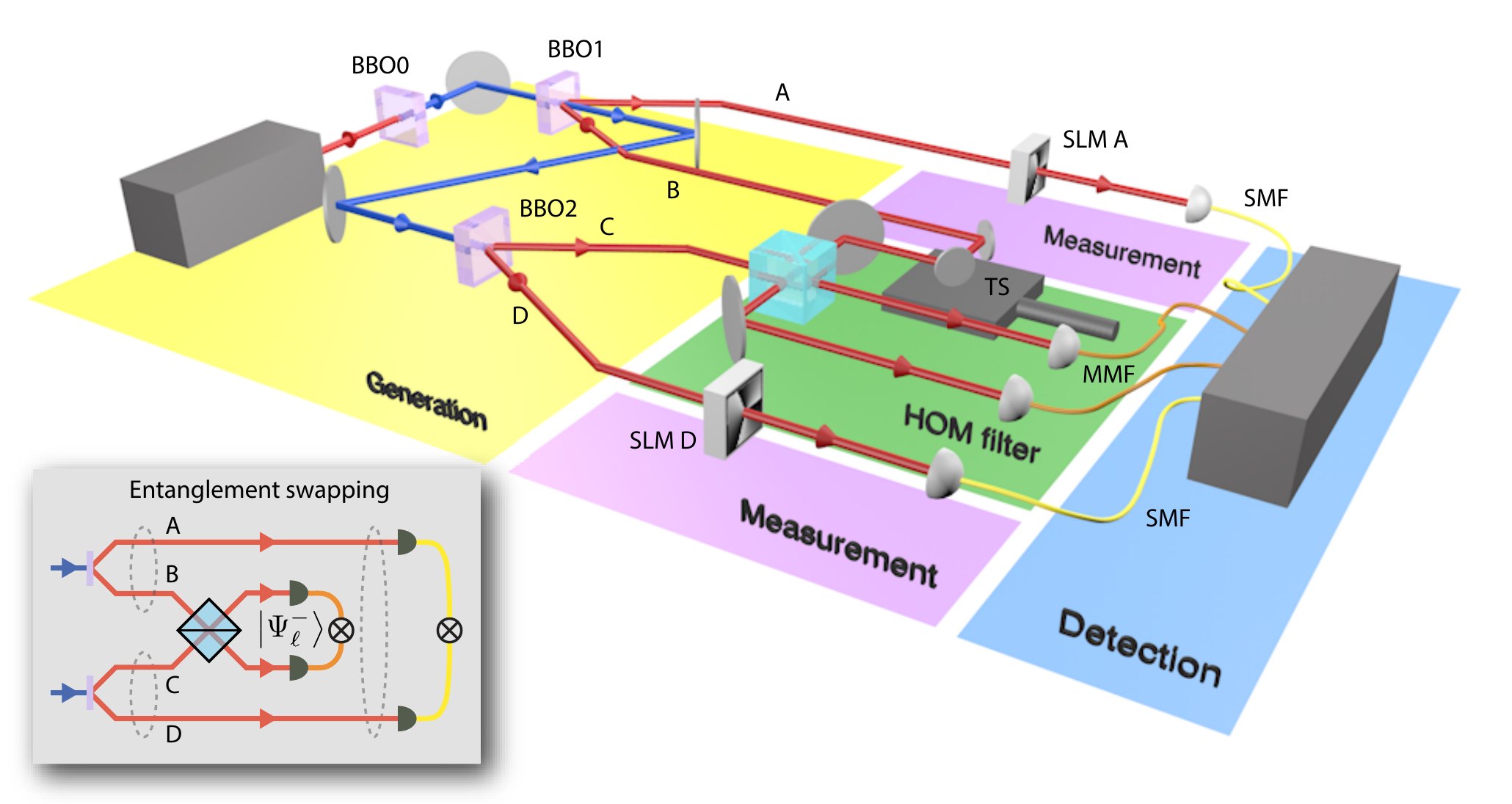}
\caption{A simplified version of the experimental setup. BBO0 is pumped by a Ti:Sapphire laser to produce UV light via upconversion. BBO1 produces a downconverted pair A and B; BBO2 produces a downconverted pair C and D. Each is entangled in the state $\ket{\Psi^+}$. The path length of B is adjusted using a translation stage (TS) such that B and C interfere on a beamsplitter; they are projected onto the antisymmetric state when detected in coincidence in the multi-mode fibres (MMFs). At this point, photons at A and D become entangled, which we measure using spatial light modulators (SLMs) in combination with single-mode fibres (SMFs). Inset: A conceptual diagram of entanglement swapping. Entanglement between A and B is transferred to A and D via interference at a beamsplitter and detection in coincidence.}
\label{setup}
\end{figure*}

In our experiment, the goal is to establish entanglement between two parties that have not interacted.  We start with two pairs of entangled photons. The first pair is an entangled state shared by Alice (A) and Bob (B); the second pair is an entangled state shared by Charlie (C) and Daisy (D).  Successful swapping corresponds to transferring the entanglement from A and B to A and D, and is equivalent to the teleportation of the state of photon B to photon D.

We generate entangled photons using SPDC in 1-mm-thick $\beta$-barium borate (BBO) crystals.  We use two crystals to generate two pairs, one in each crystal. Both crystals are pumped with $\approx$ 700 mW of light at a wavelength of 404 nm, resulting in two pairs of photons centred at 808 nm. The state at the output of each crystal is entangled in multiple degrees of freedom, including horizontal and vertical position, radial and orbital angular momentum, etc., resulting in a large multi-dimensional state with several thousand entangled modes \cite{Salakhutdinov2012,Bolduc2016}. When we consider only the OAM index, the state is given by \cite{Mair2001} 
\begin{align}\label{PDC}
\ket{\Psi}_{ij}=\sum_{\ell = 1}^\infty c_\ell \ket{\Psi_{-\ell \ell}^+}_{ij} + c_0\ket{0}_i\ket{0}_j,
\end{align}
where the squared modulus of the complex coefficients, $|c_\ell|^2$, is the probability to find both photons $i$ and $j$ in the entangled state $\ket{\Psi^{+}_{-\ell \ell}}$. The entangled state is given by $\ket{\Psi_{\ell k}^\pm}_{ij}=(\ket{\ell}_i \ket{k}_j \pm \ket{k}_i \ket{\ell}_j)/\sqrt{2}$, where $\ket{\ell}$ represents a photon with OAM $\ell\hbar$.

In the experimental aspect of this work, we focus only on the $\ell = \pm 1$ and $\ell = \pm 2$ subspaces, though the analysis can be easily extended to include the entire multi-dimensional state generated by the crystal. Considering the output of both crystals together, where the first (second) crystal produces photons A and B (C and D), we have the initial state given by
\begin{align}
\ket{\Psi} =& \left( c_1 \ket{\Psi_{-11}^+}_{\rm AB} + c_2 \ket{\Psi_{-22}^+}_{\rm AB} \right) \notag\\
& \qquad \otimes \left( c_1 \ket{\Psi_{-11}^+}_{\rm CD} + c_2 \ket{\Psi_{-22}^+}_{\rm CD} \right).
\end{align}
Note that there is no entanglement between parties A and D.  Photons B and C are then incident on a beamsplitter, where they undergo HOM interference. Our recent work \cite{Zhang2016} showed that this can act as a specific filter for the spatial modes of light, whereby any antisymmetric input state results in antibunching and a guaranteed coincidence detection. Conditioned on a coincidence between B and C, the two-photon state between photons A and D becomes
\begin{align}\label{rho}
\rho_{_{\rm AD}} = & {\cal K}^2 |c_1|^4 \ket{\Psi^{-}_{-11}}\bra{\Psi^{-}_{-11}}
+ {\cal K}^2 |c_2|^4 \ket{\Psi^{-}_{-22}}\bra{\Psi^{-}_{-22}} \nonumber \\
& + {\cal K}^2 |c_1|^2|c_2|^2 \left( \ket{\Psi^{-}_{-21}}\bra{\Psi^{-}_{-21}}
+ \ket{\Psi^{-}_{-12}} \bra{\Psi^{-}_{-12}} \right. \nonumber\\
& + \left.\ket{\Psi^{-}_{12}}\bra{\Psi^{-}_{12}}
+ \ket{\Psi^{-}_{-1-2}}\bra{\Psi^{-}_{-1-2}} \right),
\end{align}
where $\cal K$ is a normalisation factor. The result is a statistical mixture of the antisymmetric states corresponding to all combinations of two OAM values (see Supplementary Information).  Importantly, the state of A and D now contains entanglement. When we consider a particular OAM value, for example $\ell = \pm 1$, we note that the following transformation occurs
 \begin{equation}
\ket{\Psi^{+}_{-\ell \ell}}_{\rm AB} \otimes\ket{\Psi^{+}_{-\ell \ell}}_{\rm CD} \rightarrow \ket{\Psi^{-}_{-\ell \ell}}_{\rm AD} \otimes\ket{\Psi^{-}_{-\ell \ell}}_{\rm BC}\,,
\label{entswap}
\end{equation}
indicating a successful swap of entanglement from AB to AD. Note that in addition, the transformation swaps entanglement from CD to BC. However, this entanglement is lost due to the absorption of the photons BC in the detection process. 

\begin{figure}
\centering
\includegraphics[]{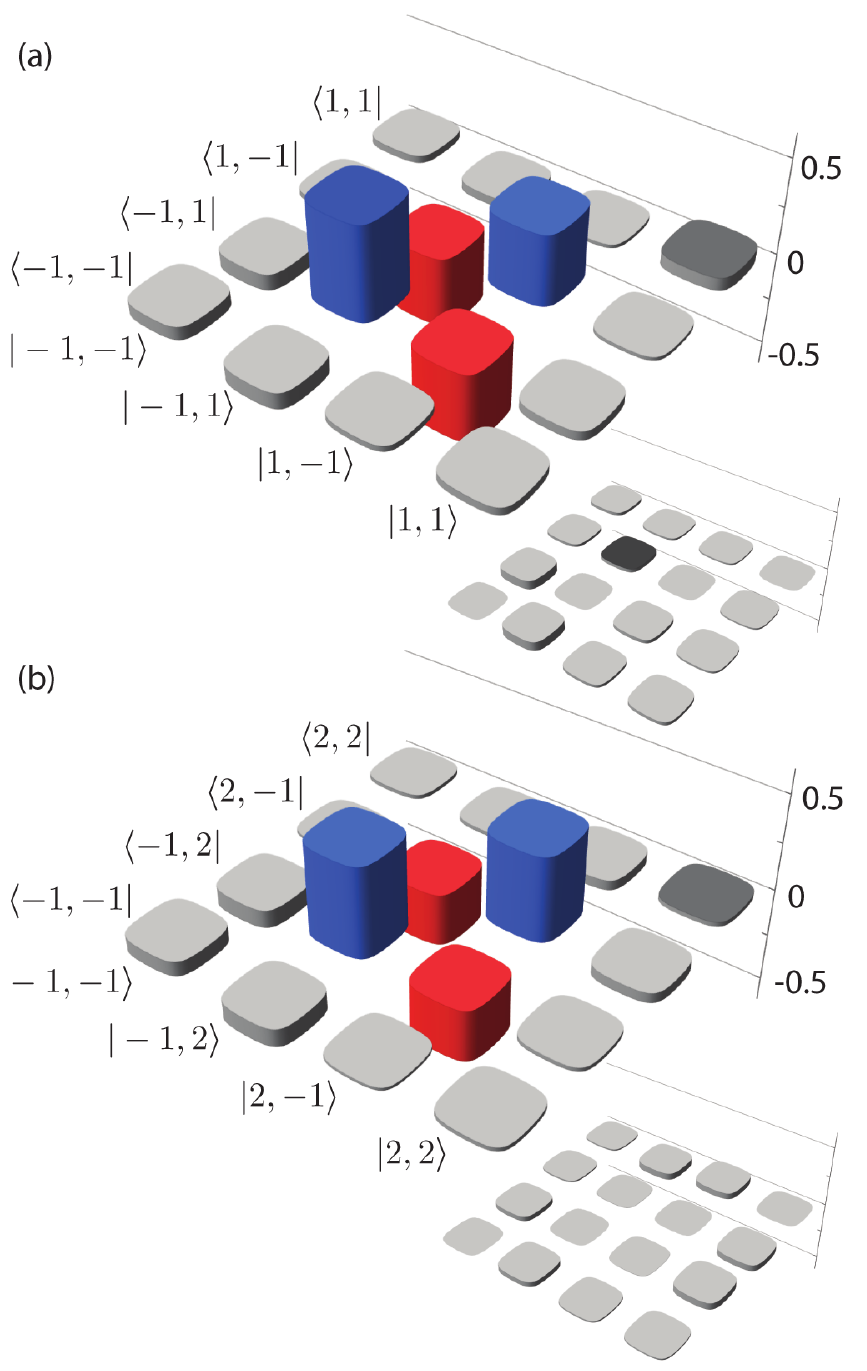}
\caption{Reconstructed density matrices of the joint state of A and D for (a) $\ell=\pm 1$ and (b) $\ell=2,-1$. Positive values are shown in blue, while negative values are shown in red; grey bars indicate the absolute value is less than 0.1. The main images show the real part of the state, while the insets show the imaginary part.}
\label{dm}
\end{figure}

We also note that entangled states are created that did not exist prior to the beamsplitter, i.e.,~$\ket{\Psi^-_{-21}}, \ket{\Psi^-_{-12}}, \ket{\Psi^-_{12}}$, and $\ket{\Psi^-_{-2-1}}$.   One can see this as the result of a transcription process where the basis for one of the subsystems in the initial state is replaced by a different basis in the final state. Transcription is commonly performed to produce OAM entanglement from polarisation entanglement by imprinting one OAM $\ell_1$ on horizontally polarised light and another OAM $\ell_2$ on vertically polarised light \cite{Fickler2012}. These transcription processes come about because of the different combinations of the terms allowed in four dimensions: the photon pair in BC (as well as the pair in AD, due to conservation of angular momentum) can be projected into one of six antisymmetric states by a coincidence detection after the beam splitter  \cite{Goyal2014, Zhang2016}.  See the Supplementary Information for further details and an example of the process.

In order to verify experimentally that our scheme successfully swaps entanglement from the photons in AB to the photons in AD, the state of photons A and D is determined using projective measurements with a combination of spatial light modulators (SLMs) and single-mode fibres (SMFs). We perform tomography on the state of photons A and D to determine the degree of entanglement. See Methods for further experimental details.

We perform full two-qubit tomography on the $\ell = \pm 1$; $\ell = \pm 2$; $\ell = 2, -1$; $\ell = 2, 1$; $\ell = 1, -2$; and $\ell = 1, 2$ subspaces. We display the reconstructed density matrices of $\ell = \pm 1$ and $\ell = 2, -1$ in Fig.~\ref{dm}, while the other four can be seen in the Supplementary Information.

The fidelity of each state with the ideal state $\ket{\Psi^-}\bra{\Psi^-}$ is an indicator of the success of the entanglement swapping. The fidelities of our two-dimensional reconstructed states are shown in Table \ref{fidcon}; they have an average fidelity of $0.80 \pm 0.05$. The maximum fidelity in our entanglement swapped states is dictated by the visibility of our HOM dip (see Methods), which is comparable to the visibility obtained in other experiments \cite{Malik2016}.

Concurrence is a convenient measure of entanglement for two dimensional subspaces (see Methods); nonzero concurrence indicates the existence of entanglement, with unit concurrence indicating maximal entanglement. We find a nonzero concurrence for all of the subspaces we reconstruct, as shown in Table \ref{fidcon}; the average is $0.7 \pm 0.1$. This nonzero concurrence indicates successful swapping in multiple two-dimensional subspaces.

\begin{figure}
\centering
\includegraphics[]{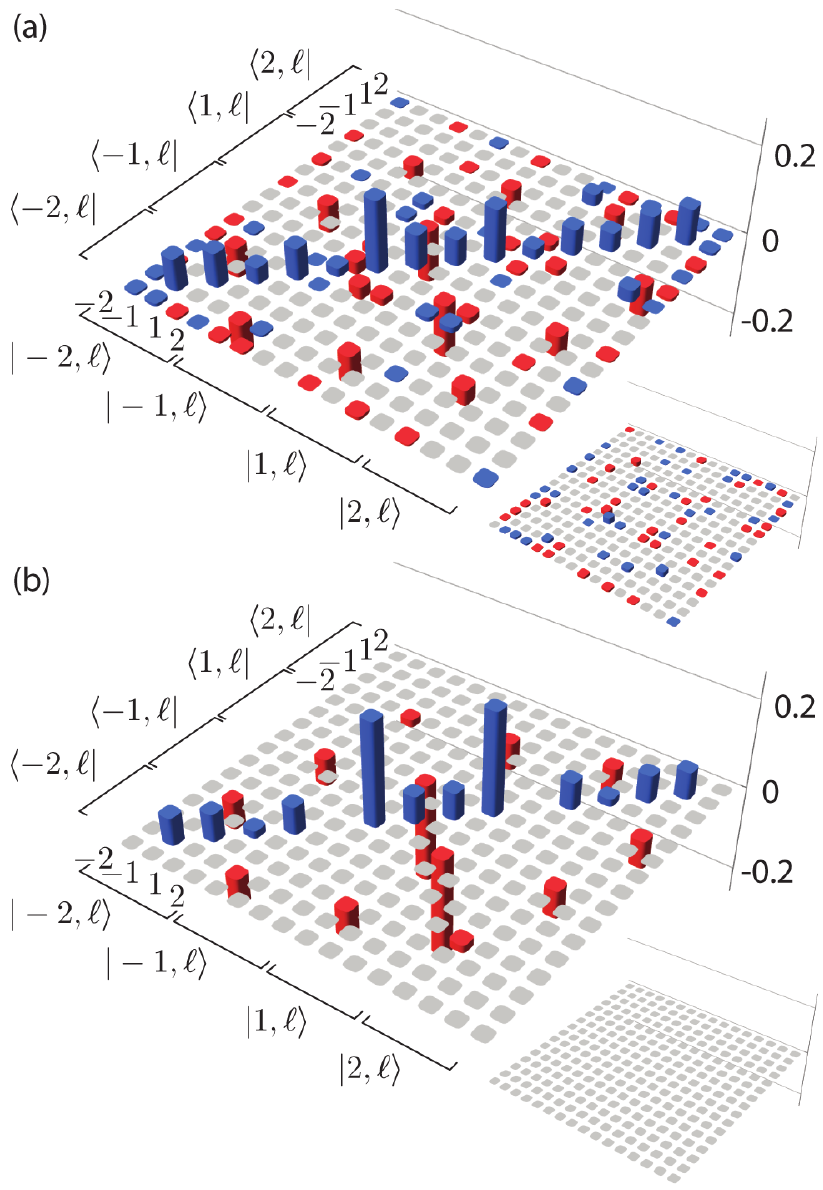}
\caption{Estimated density matrix of the joint state of A and D for the four-dimensional space with $\ell=\pm 1, \pm 2$. (a) The state estimated using the reconstructed density matrices of all six two-dimensional subspaces in Eq.~(\ref{rho}). (b) The theoretical prediction using the experimentally observed spiral bandwidth. Positive values are shown in blue, while negative values are shown in red; grey indicates the element is unmeasured in (a) or zero in (b).}
\label{fullstate}
\end{figure}

In order to estimate the four-dimensional state, we sum the density matrices of the six subspaces together according to Eq.~(\ref{rho}). The resulting state is shown in Fig.~\ref{fullstate}. The elements of the matrix that remain unmeasured are expected to be zero; these do not affect the fidelity of the final state as measurement of the fidelity requires only the diagonal elements and off-diagonal non-zero elements \cite{Malik2016}. The fidelity of our estimated state with respect to the state in Eq.~(\ref{rho}) is $0.85 \pm 0.02$, indicating a good overlap between the states.

\begin{center}
\begin{table}
\caption{Fidelity and concurrence for each of the six two-dimensional subspaces.}
\label{fidcon}
\begin{tabular}{ C{2cm} C{2cm} C{2cm} }
 \hline
 Subspace & Fidelity & Concurrence \\
 \hline
 $\ell=\pm 1$ & $0.80 \pm 0.05$ & $0.72 \pm 0.06$ \\
 $\ell=\pm 2$ & $0.85 \pm 0.05$ & $0.84 \pm 0.06$ \\
 $\ell= -2, -1$ &  $0.83 \pm 0.03$ & $0.79 \pm 0.05$ \\
 $\ell= -2, 1$ &  $0.79 \pm 0.05$ & $0.65 \pm 0.08$ \\
 $\ell= 2, -1$ &  $0.72 \pm 0.05$ & $0.57 \pm 0.08$ \\
 $\ell= 2, 1$ &  $0.82 \pm 0.02$ & $0.78 \pm 0.05$ \\
 \hline
\end{tabular}
\end{table}
\end{center}

As our work is the first demonstration of entanglement swapping for spatial states, it represents an important first step towards realising a quantum repeater for spatial modes of light.  Moreover, we note that entanglement swapping implies that teleportation has also taken place. In fact, specific single-photon teleported states have been measured in this work, and these states can be extracted from the two-photon reconstructed states shown in Fig.~\ref{dm}.

In our implementation, the final state between photons A and D is a mixture of all possible two-qubit antisymmetric entangled states $\ket{\Psi^-}$.  Considering the multi-dimensional nature of the light generated by SPDC, we estimate there to be several thousand entangled modes in this state \cite{Bolduc2016, Krenn2014}.   To experimentally measure this number of modes, one needs to take into account both the OAM and radial indices. 

For the future, in contrast to previous work \cite{Goyal2014},  it is possible to achieve a final state between photons A and D that is pure, i.e., a high-dimensionally entangled state analogous to that of Eq.~(\ref{PDC}).   An additional BBO crystal can be used to up-convert photons B and C \cite{Guerreiro2013}.  If the up-converted photon is detected in the $\ell = 0$ mode, photons A and D are projected into a pure state.   Such a mechanism generates pure high-dimensional entanglement without the need for additional photons.  Further details of this mechanism are detailed in the Supplementary Information. 

Furthermore, the present method generates entanglement between modes that were not entangled in the parent photon pairs and thus provides the ability to transcribe entanglement as explained in the Supplementary Information. We believe that the correlations between photons that have not interacted with each other will find applications in remote state engineering, remote ghost imaging, and multi-party quantum key distribution.

In conclusion, we have demonstrated for the first time entanglement swapping of OAM states of light. We have confirmed the completion of the entanglement swapping by performing complete tomography of the final entangled pair in multiple two-dimensional OAM subspaces.   For all of the subspaces that we consider, we measure a final concurrence greater than zero, indicating that our swap was successful.  This result confirms that we have achieved entanglement swapping for multiple OAM subspaces.  For each subspace, we obtained an average fidelity of 80\% between the reconstructed state and the maximally entangled antisymmetric state.  This can be viewed as the first step to building a quantum repeater with spatial modes of light, an essential ingredient for broadband long-distance quantum communication.


%

\section{Methods}

\begin{figure*}
\centering
\includegraphics[]{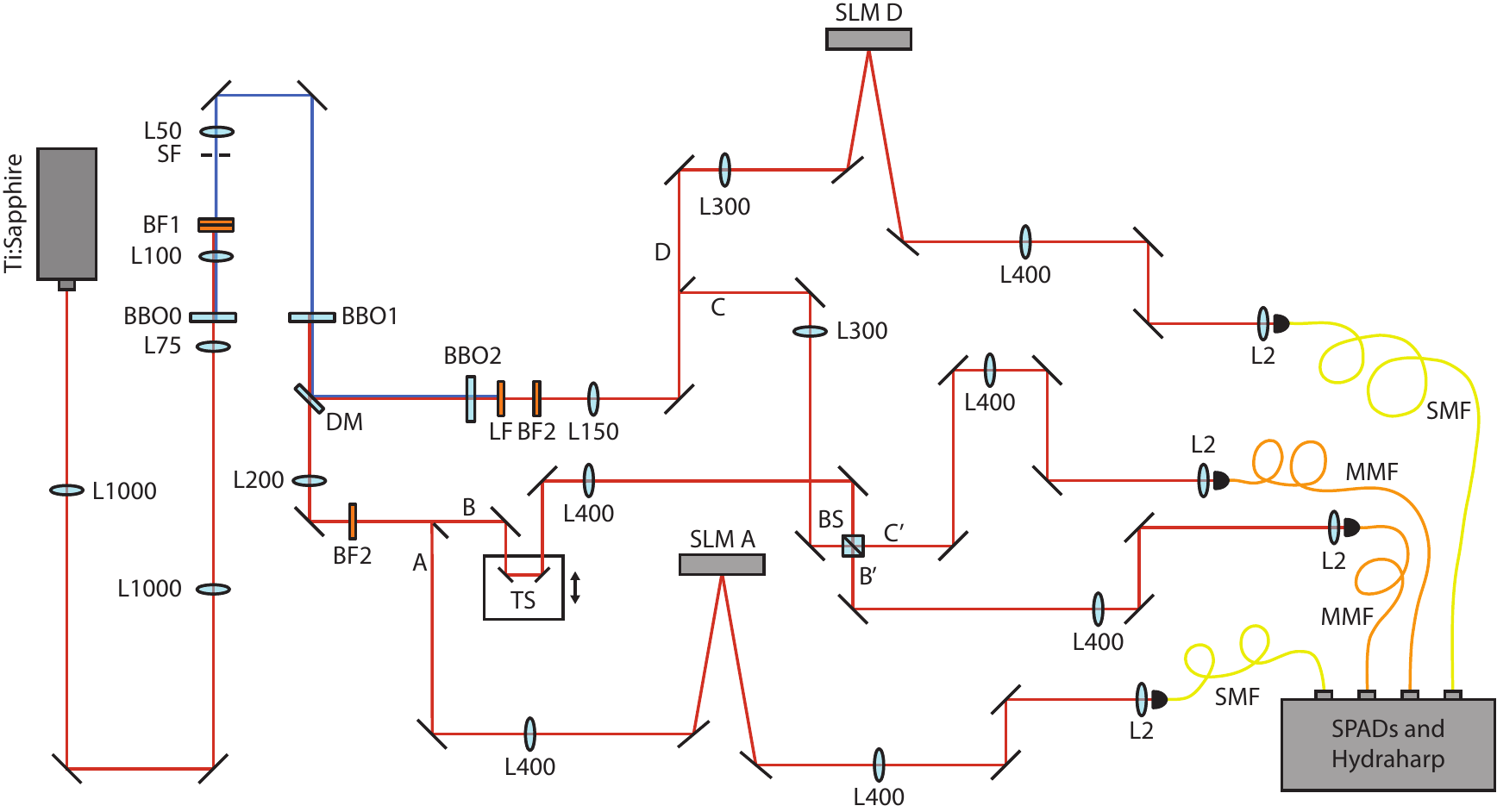}
\caption{A detailed schematic of the experimental setup. Abbreviations can be found in the Methods section.}
\label{detailedsetup}
\end{figure*}

\subsection{Experimental Details}

As seen in Fig.~\ref{detailedsetup}, our experiment uses a pulsed Ti:sapphire laser (Coherent Chameleon Ultra II) centred at 808 nm, with a pulse width of 140 fs and a repetition rate of 80 MHz. We image the output plane of the laser to the beginning of our setup using two lenses of focal length 1000 mm (L1000). Using a lens of focal length 75 mm (L75), we focus the laser into a 0.5-mm-thick $\beta$-barium borate (BBO) crystal (BBO0). The resultant sum frequency generation produces $\approx$ 700 mW of ultraviolet (UV) light at 404 nm. We focus the upconverted light through a 100-$\mu$m circular aperture (spatial filter SF) using a 100-mm lens. The light that passes through the aperture is collimated with a 50-mm lens. The spatial filtering at the aperture ensures that the pump beam used for the downconversion has a Gaussian beam profile. The remaining infrared light is removed using two consecutive bandpass filters BF1 (10-nm width centred at 405 nm).

The UV light is used to pump a 1-mm BBO crystal (BBO1), producing pairs of photons at 808 nm via type-I, near-collinear SPDC. The remaining UV light is deflected using a dichroic mirror, and the downconverted light continues on through lens L200 and bandpass filter BF2 (3-nm width centred at 808 nm). It is then split using a D-shaped mirror so that one photon continues on as photon B and the other is reflected as photon A.

Photon B strikes two mirrors on a motorised translation stage (TS) for precise path length adjustment. Photon B then passes through a 400-mm lens (L400) before striking a non-polarising beamsplitter (BS) in the image plane of BBO1. Meanwhile, photon A passes through L400 before striking SLM A in the image plane of BBO1. SLM A is imaged to a single-mode fibre (SMF) using L400 and a 2-mm lens (L2).

After being deflected by the dichroic mirror, the UV light then pumps a second 1-mm BBO crystal (BBO2), after which it is filtered out using longpass filter LF (cutoff wavelength 750 nm). A second pair of photons at 808 nm is produced via SPDC and passes through lens L150 and bandpass filter BF2. It is then split with a D-shaped mirror so that one photon continues on as photon D and the other is reflected as photon C.

Photon D passes through a 300-mm lens (L300) before striking SLM D in the image plane of BBO2. SLM D is imaged to an SMF using L400 and L2. Photon C passes through L300 before striking the BS in the image plane of BBO2.

\begin{figure}
\centering
\includegraphics[]{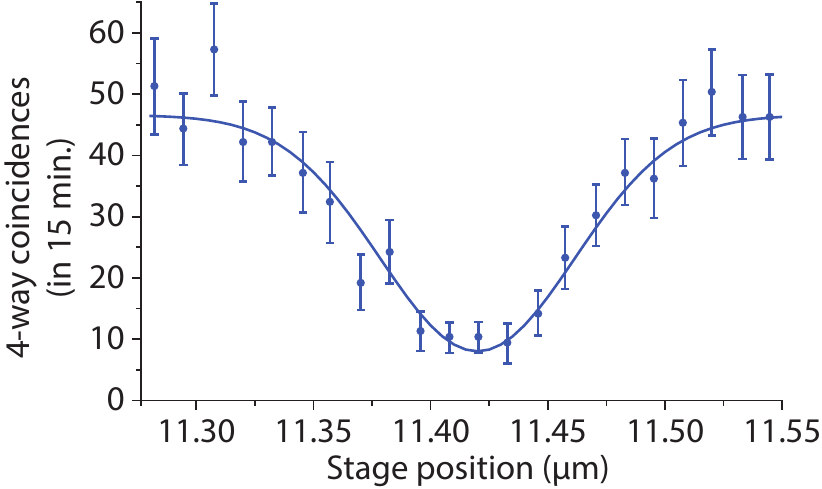}
\caption{Hong-Ou-Mandel interference of photons in the state $\ell=0$. We fit a Gaussian curve centred at 11.42 $\mu$m with a visibility of 0.71. Error bars correspond to Poisson statistics in the count rate.}
\label{hom}
\end{figure}

Here photons B and C undergo Hong-Ou-Mandel (HOM) interference; the exact position of the HOM interference dip is identified by moving the translation stage in path B until a drop in the four-photon coincidence rate is observed. This HOM dip is shown in Fig.~\ref{hom}. After the BS, the new paths B$'$ and C$'$ are each imaged to multi-mode fibres (MMFs) using L400 and L2.

Each of the four fibres is connected to a single-photon avalanche detector (SPAD, Excelitas SPCM-800-14-FC), which is in turn connected to a coincidence detection system (HydraHarp). The average count rate for the $\ell= \pm 1$ subspace is 0.04 Hz, while the average count rate for the $\ell= \pm 2$ subspace is 0.01 Hz.

The combined two-dimensional state of photons A and D is determined by displaying holograms of four OAM states on each SLM in turn: $\ket{\ell_1}$, $\ket{\ell_2}$, $(\ket{\ell_1}+\ket{\ell_2})/\sqrt{2}$, and $(\ket{\ell_1}+i\ket{\ell_2})/\sqrt{2}$. Using the 16 resulting measurements, we reconstruct the density matrix using quantum state tomography.

\subsection{Spatial light modulators}

The OAM of light can be measured with the combination of a spatial light modulator and a single-mode fibre. An SLM displays a computer-generated hologram of an OAM mode $\ell_{SLM}$; the phase displayed by the SLM is added to that of the incident light. Any reflected light with OAM $\ell=0$ will successfully couple into the fibre. The detected light then must have had OAM $\ell=-\ell_{SLM}$ prior to striking the SLM.

\subsection{Fidelity vs. Visibility}

The fidelity of a density matrix $\rho$ with another density matrix $\sigma$ is
\begin{equation}
F={\rm Tr}\left(\sqrt{\sqrt{\rho} \sigma \sqrt{\rho}}\right)^2.
\end{equation}
Unit fidelity indicates perfect overlap between the states, while zero fidelity indicates no overlap between the states.

The visibility of the HOM dip limits the quality of results. With a visibility of $V$, the entanglement swapping only occurs $V\%$ of the time. Then $(1-V)\%$ of the time, the interference at the beamsplitter is unsuccessful, and the resultant four-way coincidences represent uncorrelated noise. Under this assumption, the total two-dimensional state measured is then given by
\begin{equation}
\rho_{\rm AD} = V \ket{\Psi^-} \bra{\Psi^-} + (1-V) \frac{\mathbb{I}}{4},
\end{equation}
where $\mathbb{I}$ is the identity matrix.

\begin{figure}
\centering
\includegraphics[]{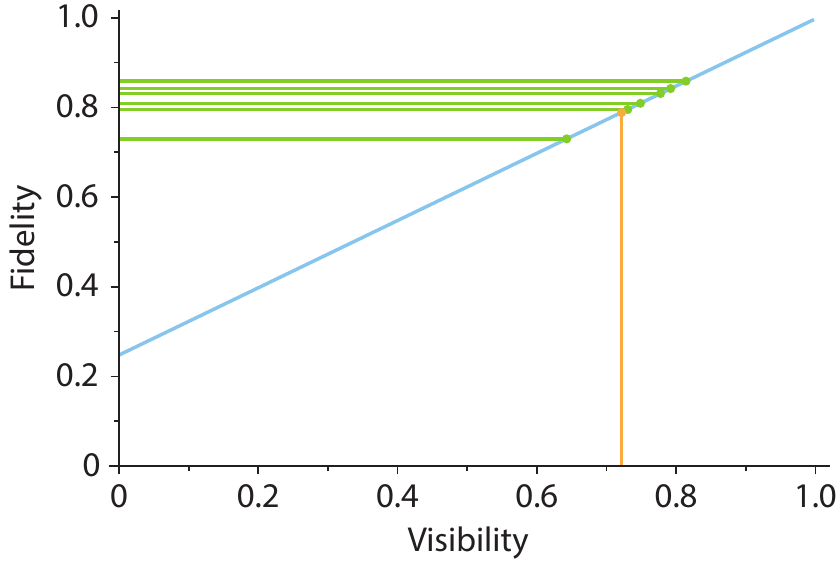}
\caption{Fidelity of the predicted state with the ideal state as a function of visibility. Green points correspond to the fidelities of the six two-dimensional subspaces and have uncertainties as shown in table \ref{fidcon}. The orange point corresponds to the measured four-way visibility.}
\label{fidvis}
\end{figure}

The fidelity of the predicted state $\rho_{\rm AD}$ with the ideal state $\ket{\Psi^-} \bra{\Psi^-}$ as a function of visibility is shown in Fig.~\ref{fidvis}. A visibility of approximately 71\% as in our experiment produces a fidelity of 78\%, which is consistent with our results.

\subsection{Concurrence}

The concurrence of a density matrix $\rho$ is calculated by first obtaining a matrix
\begin{equation}
R=\sqrt{\sqrt{\rho}\,\widetilde{\rho}\,\sqrt{\rho}\,},
\end{equation}
where $\widetilde{\rho}=(\sigma_y \otimes \sigma_y) \rho^* (\sigma_y \otimes \sigma_y)$. Here $\sigma_y$ represents the Pauli spin matrix and $\rho^*$ represents the complex conjugate of $\rho$. The eigenvalues of the matrix $R$ are denoted $\lambda_1,\lambda_2,\lambda_3,\lambda_4$ in decreasing order. Then the concurrence of $\rho$ is
\begin{equation}
C(\rho)={\rm max}(0,\lambda_1-\lambda_2-\lambda_3-\lambda_4).
\end{equation}
Nonzero concurrence indicates the state is entangled. Unit concurrence indicates a maximally entangled state.

\subsection{Background subtraction}

\subsubsection{1. Expected 4-way coincidence}

Consider a laser with a repetition rate of $R$ pumping a nonlinear crystal to generate an entangled photon pair A and B via SPDC. If $C_{\rm AB}$ is the number of coincidence events per second detected between detectors A and B, then $C_{\rm AB}/R$ is the probability that a coincidence event will be detected (or generated\footnote{The probability of detection and generation differ only by a constant factor, the detection efficiency, so any expression worked out for the one should be proportional to the other.}) from a single laser pulse.

Now consider the same laser pulse pumping a second crystal to generate a second pair of photons. The probability to detect/generate these two uncorrelated photon pairs from the same laser pulse, with one pair detected at detectors A and B and the other at C and D, is given by
\begin{equation}
\frac{C_{\rm AB}C_{\rm CD}}{R^2}.
\label{prob}
\end{equation}
Then the rate per second is given by Eq.~(\ref{prob}) multiplied by the repetition rate $R$, which gives
\begin{equation}
\frac{C_{AB}C_{CD}}{R}.
\end{equation}

In our experiment a photon pair generated by BBO1 can be detected in coincidence by detectors A\&B or A\&C, and a second pair generated by BBO2 can be detected by detectors B\&D or C\&D, so we add all the combinations that can result in coincidence between all 4 detectors. Therefore the number of 4-way coincidence events per second generated by two entangled photon pairs is
\begin{equation}
\frac{1}{R}\left(C_{\rm AB}C_{\rm CD} + C_{\rm AC}C_{\rm BD}\right).
\label{rate}
\end{equation}

\subsubsection{2. Background of 4-way coincidence}

There are two ways in which 4-way coincidence events can be generated from uncorrelated/non-entangled photons. Firstly, a 2-way coincidence event from entangled photons can be detected together with two uncorrelated single detection events. Secondly, four uncorrelated single events can be detected in coincidence. Thus from Eq.~(\ref{rate}) we find the expected number of background events per second to be
\begin{align}
\frac{1}{R^2}&\left(C_{AB}S_{C}S_{D} + S_{A}S_{B}C_{CD} + C_{AC}S_{B}S_{D} + S_{A}S_{C}C_{BD}\right) \notag\\ +&\frac{1}{R^3}\left(S_{A}S_{B}S_{C}S_{D}\right),
\end{align}
where $S_i$ is the number of single events per second detected in detector $i$. We subtract the calculated number of background counts from the measured data to obtain the actual number of counts. Occasionally with count rates that are expected to be very low, the measured number of counts is smaller than the expected number of background counts; in this case, we replace the count rate with zero.

\section{Author Contributions}

YW, MA, TR, and JL performed the experiment. YW and MA performed the data analysis. FSR and TK provided the theoretical framework. MA wrote the first draft of the paper, and all authors contributed to the final version of the manuscript. JL, FSR, AF, and DF supervised the project, and the idea was conceived by JL, FSR, AF and TK.

\section{Acknowledgements}

We acknowledge support from the European Research Council under the European Union's Seventh Framework Programme (FP/2007-2013)/ERC GA 306559, the Engineering and Physical Sciences Research Council (EPSRC, UK, grants EP/M006514/1, EP/M01326X/1), the Photonics Initiative of South Africa, and the Natural Sciences and Engineering Research Council of Canada (NSERC).

\clearpage

\section{Supplementary Information: Entanglement Swapping for Orbital Angular Momentum Modes}

The state of the two photon pairs produced by spontaneous parametric downconversion (SPDC) in paths A, B and C, D, respectively, can be written as
\begin{align}
\ket{\psi_0}= &\left(c_0 \kets{0}{\rm A}\!\kets{0}{\rm B} + \sum_n c_{n} \kets{\Psi^{+}_{n\overline{n}}}{\rm AB} \right) \nonumber \\
&\otimes \left( c_0 \kets{0}{\rm C}\!\kets{0}{\rm D} + \sum_m c_{m} \kets{\Psi^{+}_{m\overline{m}}}{\rm CD }\right) ,
\label{initial}
\end{align}
where $c_n$ represents complex coefficients, we use the notation $\overline{n}:=-n$ and
\begin{equation}
\ket{\Psi^{\pm}_{\ell \ell'}}:= \frac{1}{\sqrt{2}}\left(\ket{\ell}\ket{\ell'} \pm \ket{\ell'}\ket{\ell}\right) ,
\end{equation}
denote symmetric and antisymmetric Bell states with a plus and a minus sign, respectively.

The photons in path B and path C are subjected to a $50:50$ beam splitter which superposes the two beams. The inversion of the helicity of the OAM modes, $\ell \rightarrow -\ell$, upon reflection in the beam spitter is compensated by two additional reflections employing a mirror in path B before and another one in path C behind the beam splitter (cp. Fig.\ 1 in the main text). The action of the beam splitter, combined with both mirrors, is thus characterised by the transformation rules\footnote{We denote the original before the beam splitter and the reflected path thereafter by the same letter. Accordingly a photon in the input port B exits in path B upon reflection and in path C upon transmission.}
\begin{eqnarray}
\kets{\ell}{\rm B}& \rightarrow & \frac{1}{\sqrt{2} }\left(\kets{\ell}{\rm C} - \kets{\ell}{\rm B}\right) \label{bs1}\\
\kets{\ell}{\rm C}& \rightarrow & \frac{1}{\sqrt{2} }\left(\kets{\ell}{\rm B} + \kets{\ell}{\rm C} \right)
\label{bs2}
\end{eqnarray}
Using these transformation rules on the input state in (\ref{initial}), we obtain a state after the beam splitter, under the condition that each output path of the beam splitter contains a single photon,\footnote{The photons in paths B and C are detected in coincidence.} that only consists of antisymmetric photon pairs. For $d$-dimensions, one can expression it as
\begin{widetext}
\begin{align}
\ket{\psi_1} = & {\cal K} \left[ \sum_{n=1}^N c_{n}^2 \kets{\Psi^{-}_{n\overline{n}}}{\rm AD} \kets{\Psi^{-}_{n\overline{n}}}{\rm BC} - \sum_{n=1}^N c_0 c_{n} \left( \kets{\Psi^{-}_{0n}}{\rm AD} \kets{\Psi^{-}_{0\overline{n}}}{\rm BC} + \kets{\Psi^{-}_{0\overline{n}}}{\rm AD} \kets{\Psi^{-}_{0n}}{\rm BC} \right) \right. \nonumber \\
& - \left. \sum_{m\neq n=1}^N c_{m}c_{n} \left( \kets{\Psi^{-}_{n\overline{m}}}{\rm AD} \kets{\Psi^{-}_{\overline{n}m}}{\rm BC}
 + \kets{\Psi^{-}_{\overline{n}m}}{\rm AD} \kets{\Psi^{-}_{n\overline{m}}}{\rm BC} + \kets{\Psi^{-}_{nm}}{\rm AD} \kets{\Psi^{-}_{\overline{n}\overline{m}}}{\rm BC}
 + \kets{\Psi^{-}_{\overline{n}\overline{m}}}{\rm AD} \kets{\Psi^{-}_{nm}}{\rm BC} \right) \right] ,
\label{combo}
\end{align}
\end{widetext}
where ${\cal K}$ is a normalization constant that compensates for the loss of the terms with two photons in the same output path of the beam splitter and $d=1+2N$. In the case where we only consider $\ell=\pm 1,\pm 2$, we have
\begin{align}
&\ket{\psi_1} = \nonumber \\
& {\cal K} \left[ c_1^2 \kets{\Psi^{-}_{1\overline{1}}}{\rm AD} \kets{\Psi^{-}_{1\overline{1}}}{\rm BC} + c_2^2 \kets{\Psi^{-}_{2\overline{2}}}{\rm AD} \kets{\Psi^{-}_{2\overline{2}}}{\rm BC} \right. \nonumber  \\
& - c_1 c_2 \left( \kets{\Psi^{-}_{1\overline{2}}}{\rm AD} \kets{\Psi^{-}_{\overline{1}2}}{\rm BC} + \kets{\Psi^{-}_{\overline{1}2}}{\rm AD} \kets{\Psi^{-}_{1\overline{2}}}{\rm BC} \right. \nonumber\\
& \left. + \left.\kets{\Psi^{-}_{12}}{\rm AD} \kets{\Psi^{-}_{\overline{1}\overline{2}}}{\rm BC} + \kets{\Psi^{-}_{\overline{1}\overline{2}}}{\rm AD} \kets{\Psi^{-}_{12}}{\rm BC} \right) \right].
\end{align}

Simultaneous detection of a single photon in each of the two output ports of a symmetric beam splitter causes a projection onto the antisymmetric component of the input state (cp.\ Ref [18] in the main text). The dimension of the corresponding antisymmetric state space is given by the number of ways in which the OAM values of the input space can be combined into pairs of $\ell$'s to form an antisymmetric state $\ket{\Psi^{-}_{\ell \ell'}}$. For example, four OAM values would give six antisymmetric basis states, featuring in the BC-components of the state in (\ref{combo}). On the other hand, the antisymmetric basis states also feature as components of the photon pair in AD, because the OAM must sum to zero in each term of the state. In general, considering $d$ OAM levels, we can produce a state of the form given in (\ref{combo}) with our setup consisting of $d(d-1)/2$ antisymmetric basis states that involve both photon pairs.

Note that the state expressed in (\ref{combo}) represents the Schmidt decomposition of an entangled state, i.e., it has the form
\begin{equation}
\ket{\psi_1} = \sum_i c_i \kets{\phi_i}{\rm AD}\kets{\phi_i}{\rm BC} ,
\end{equation}
where the Schmidt bases are photon pairs in AD and BC, respectively. So, apart from the entanglement among the different pairs, there is also the maximal entanglement within the pairs between the single photons.

The detection of the photons in paths B and C without measurement of their OAM values, results in a statistical mixture of the antisymmetric states obtained from the state in (\ref{combo}). By tracing over the OAM degrees of freedom of the photons in paths B and C, we obtain for $d$-dimensions
\begin{widetext}
\begin{align}
\rho_{_{\rm AD}} = & {\cal K}^2 \left[ \sum_{n=1}^N |c_{n}|^4 \ket{\Psi^{-}_{n\overline{n}}} \bra{\Psi^{-}_{n\overline{n}}}
+ \sum_{n=1}^N |c_0|^2 |c_{n}|^2 \left( \ket{\Psi^{-}_{0n}} \bra{\Psi^{-}_{0n}} + \ket{\Psi^{-}_{0\overline{n}}} \bra{\Psi^{-}_{0\overline{n}}} \right) \right. \nonumber \\
& + \left. \sum_{m\neq n=1}^N |c_{m}|^2 |c_{n}|^2 \left( \ket{\Psi^{-}_{n\overline{m}}} \bra{\Psi^{-}_{n\overline{m}}}
 + \ket{\Psi^{-}_{\overline{n}m}} \bra{\Psi^{-}_{\overline{n}m}} + \ket{\Psi^{-}_{nm}} \bra{\Psi^{-}_{nm}}
 + \ket{\Psi^{-}_{\overline{n}\overline{m}}} \bra{\Psi^{-}_{\overline{n}\overline{m}}} \right) \right] .
\label{combo0}
\end{align}
\end{widetext}
Restricted to $\ell=\pm 1,\pm 2$, the result reduces to
\begin{align}
\rho_{_{\rm AD}} = & {\cal K}^2 |c_1|^4 \ket{\Psi^{-}_{1\overline{1}}}\bra{\Psi^{-}_{1\overline{1}}}
+ {\cal K}^2 |c_2|^4 \ket{\Psi^{-}_{2\overline{2}}}\bra{\Psi^{-}_{2\overline{2}}} \nonumber \\
& + {\cal K}^2 |c_1|^2|c_2|^2 \left( \ket{\Psi^{-}_{1\overline{2}}}\bra{\Psi^{-}_{1\overline{2}}}
+ \ket{\Psi^{-}_{\overline{1}2}} \bra{\Psi^{-}_{\overline{1}2}} \right. \nonumber\\
& + \left.\ket{\Psi^{-}_{12}}\bra{\Psi^{-}_{12}}
+ \ket{\Psi^{-}_{\overline{1}\overline{2}}}\bra{\Psi^{-}_{\overline{1}\overline{2}}} \right) .
\end{align}
The projection onto the antisymmetric space of the photons in B and C, transfers entanglement between the systems in A and B to the remote systems in A and D, which were not entangled before. This constitutes entanglement swapping.

We note that, by using a filter in paths A and D that projects onto any two-dimensional subspace with OAM values $\{\ell, \ell'\}$, one obtains an antisymmetric state $\keti{\Psi^{-}_{\ell,\ell'}}_{_{\rm AD}}$, which is maximally entangled. Such a filter in front of the detectors in paths B and C could be used to prepare a particular antisymmetric state remotely in paths A and D. A similar procedure could be exploited for various purposes of quantum communication between three or four parties, such as secure bit commitment or QKD protocols.

\begin{figure*}
\centering
\includegraphics[]{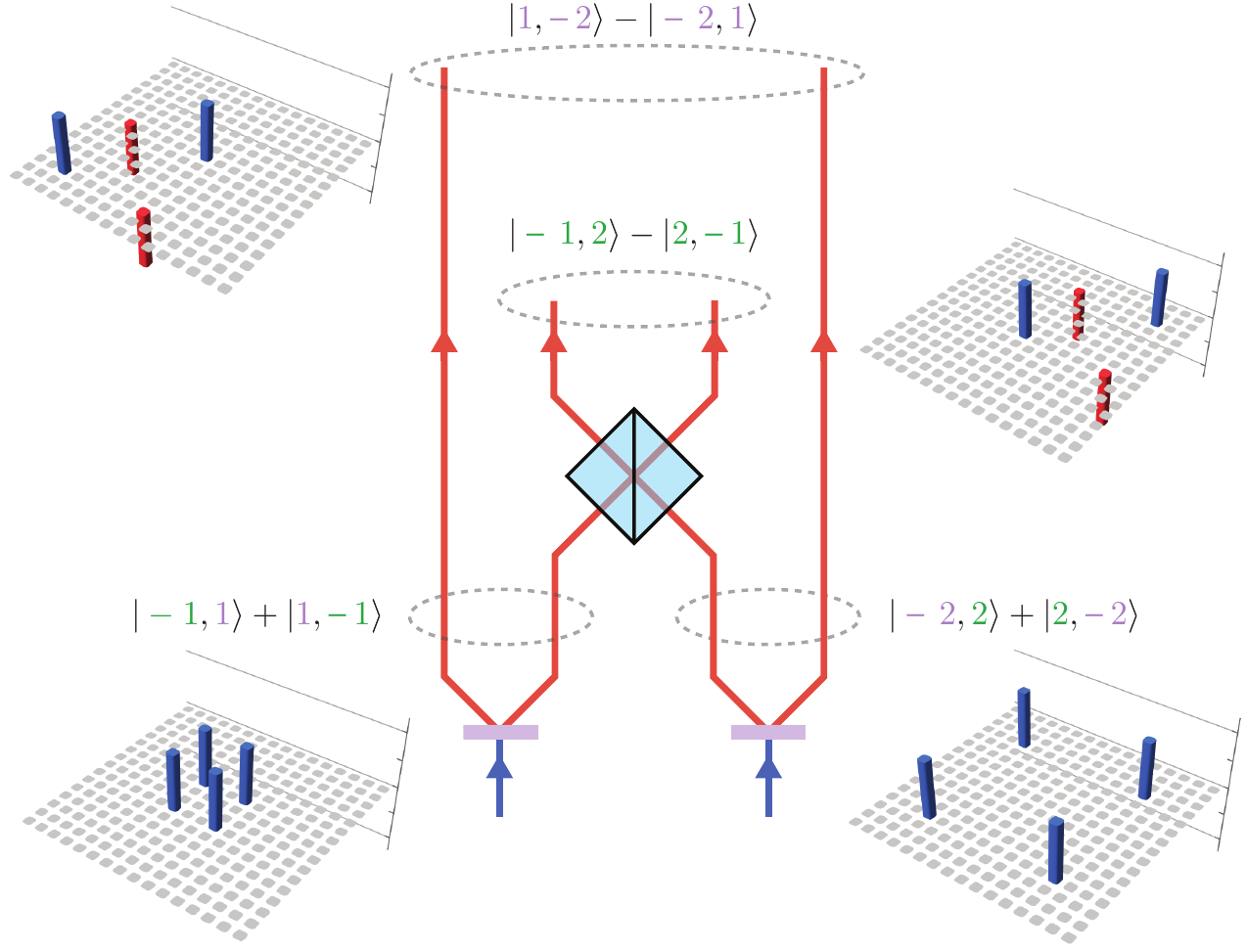}
\caption{Example of the transcription process. We start with the entangled states $\ket{\Psi^+_{-11}}_{AB}$ and $\ket{\Psi^+_{-22}}_{CD}$. After the beamsplitter and projection onto the appropriate anti-symmetric state  $\ket{\Psi^-_{-12}}$, the state between photons A and D is $\ket{\Psi^-_{-21}}$. The OAM values shown in green become the OAM values in the state projected onto B and C. The OAM values shown in purple become the OAM values in the state between photons A and D.  States shown without normalisation for clarity. The density matrices show the corresponding maximally entangled states.}
\label{trans}
\end{figure*}

\begin{figure*}
\centering
\includegraphics[]{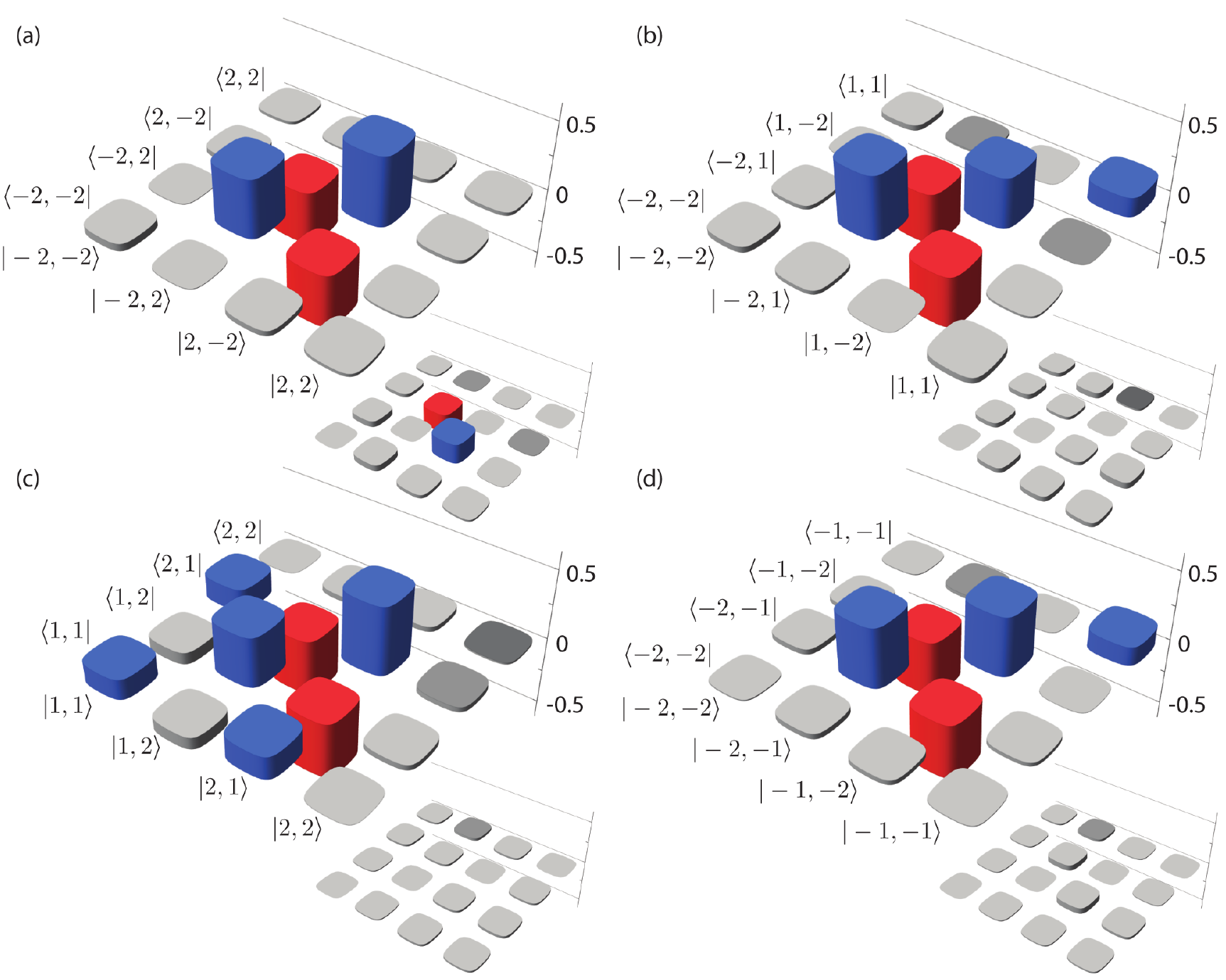}
\caption{Reconstructed density matrices of the joint state of A and D for (a) $\ell=\pm 2$, (b) $\ell=-2,1$, (c) $\ell=2,1$, and (d) $\ell=-2,-1$. Positive values are shown in blue, while negative values are shown in red; grey bars indicate the absolute value is less than 0.1. The main images show the real part of the state, while the insets show the imaginary part.}
\label{dms}
\end{figure*}

By means of particular filters for photons in BC, it is also possible to obtain a pure state with a multitude of entangled levels instead of a mixture in AD. For example, projecting on a superposition of singlet states $ \ket{\Psi^{-}_{n\overline{n}}}$ in BC results, as shown below, in a superposition of such states in AD. According to Eq.~(\ref{combo}), the state of photons in BC after the beamsplitter reads 
\begin{equation}
\ket{\Psi_1}=\sum_{n=1}^\infty\, \alpha_n\ket{n}_{_{ \rm AD}}\otimes \ket{n}_{{\rm BC}}+ \ldots
\end{equation}
with $\ket{n}\equiv\ket{\Psi^{-}_{n\overline{n}}}$ and $\alpha_n\equiv {\cal K} c_n^2$, where  components $\ket{\Psi^{-}_{n,m}}$ with different OAM values $n\not=m$ are not mentioned explicitly. A filter in BC projecting onto the state $\ket{x}\equiv (\sum_{n=1}^N \ket{n})/\sqrt{N}$ leads to 
\begin{align}
\ket{\Psi_1}\rightarrow& \left(\mathbbm{1}_{ \rm AD} \otimes \ket{x}_ {\rm BC}\bra{x}  \right)\ket{\Psi_1}\nonumber \\
& =\frac{1}{\sqrt{N}}\sum_{n=1}^\infty \sum_{m=1}^N\alpha_n \ket{n}_{{\rm AD}}\otimes \ket{x}_{ \rm BC}\bra{m}n\rangle_{\rm BC}\nonumber \\
&=\left(\frac{1}{\sqrt{N}} \sum_{n=1}^N \alpha_n  \ket{n}_{ \rm AD}\right)\otimes \ket{x}_{{\rm BC}}.
\end{align}
The resulting state of the photons in AD,  $\sum_{n=1}^N \tilde{\alpha}_n  \ket{n} \equiv \sum \tilde{\alpha}_n\ket{\Psi^{-}_{n\overline{n}}}$ (with normalised coefficients $\tilde{\alpha}_n=  c_n^2/\sqrt{\sum_n |c_n^2|^2}$), is a pure entangled state of Schmidt rank $N$. Such a filter could be realised, e.g., by parametric up-conversion of the photon pair in AD to a photon of double the frequency (the inverse process to SPDC)  and subsequent measurement of its OAM, conditioning on the OAM value $\ell=0$.


\begin{thebibliography}{33}%
\makeatletter
\providecommand \@ifxundefined [1]{%
 \@ifx{#1\undefined}
}%
\providecommand \@ifnum [1]{%
 \ifnum #1\expandafter \@firstoftwo
 \else \expandafter \@secondoftwo
 \fi
}%
\providecommand \@ifx [1]{%
 \ifx #1\expandafter \@firstoftwo
 \else \expandafter \@secondoftwo
 \fi
}%
\providecommand \natexlab [1]{#1}%
\providecommand \enquote  [1]{``#1''}%
\providecommand \bibnamefont  [1]{#1}%
\providecommand \bibfnamefont [1]{#1}%
\providecommand \citenamefont [1]{#1}%
\providecommand \href@noop [0]{\@secondoftwo}%
\providecommand \href [0]{\begingroup \@sanitize@url \@href}%
\providecommand \@href[1]{\@@startlink{#1}\@@href}%
\providecommand \@@href[1]{\endgroup#1\@@endlink}%
\providecommand \@sanitize@url [0]{\catcode `\\12\catcode `\$12\catcode
  `\&12\catcode `\#12\catcode `\^12\catcode `\_12\catcode `\%12\relax}%
\providecommand \@@startlink[1]{}%
\providecommand \@@endlink[0]{}%
\providecommand \url  [0]{\begingroup\@sanitize@url \@url }%
\providecommand \@url [1]{\endgroup\@href {#1}{\urlprefix }}%
\providecommand \urlprefix  [0]{URL }%
\providecommand \Eprint [0]{\href }%
\providecommand \doibase [0]{http://dx.doi.org/}%
\providecommand \selectlanguage [0]{\@gobble}%
\providecommand \bibinfo  [0]{\@secondoftwo}%
\providecommand \bibfield  [0]{\@secondoftwo}%
\providecommand \translation [1]{[#1]}%
\providecommand \BibitemOpen [0]{}%
\providecommand \bibitemStop [0]{}%
\providecommand \bibitemNoStop [0]{.\EOS\space}%
\providecommand \EOS [0]{\spacefactor3000\relax}%
\providecommand \BibitemShut  [1]{\csname bibitem#1\endcsname}%
\let\auto@bib@innerbib\@empty
\bibitem [{\citenamefont {Briegel}\ \emph {et~al.}(1998)\citenamefont
  {Briegel}, \citenamefont {D{\"u}r}, \citenamefont {Cirac},\ and\
  \citenamefont {Zoller}}]{Briegel1998}%
  \BibitemOpen
  \bibfield  {author} {\bibinfo {author} {\bibfnamefont {H.~J.}\ \bibnamefont
  {Briegel}}, \bibinfo {author} {\bibfnamefont {W.}~\bibnamefont {D{\"u}r}},
  \bibinfo {author} {\bibfnamefont {J.~I.}\ \bibnamefont {Cirac}}, \ and\
  \bibinfo {author} {\bibfnamefont {P.}~\bibnamefont {Zoller}},\ }\href@noop {}
  {\bibfield  {journal} {\bibinfo  {journal} {Physical Review Letters}\
  }\textbf {\bibinfo {volume} {81}},\ \bibinfo {pages} {5932} (\bibinfo {year}
  {1998})}\BibitemShut {NoStop}%
\bibitem [{\citenamefont {Pan}\ \emph {et~al.}(1998)\citenamefont {Pan},
  \citenamefont {Bouwmeester}, \citenamefont {Weinfurter},\ and\ \citenamefont
  {Zeilinger}}]{Pan1998}%
  \BibitemOpen
  \bibfield  {author} {\bibinfo {author} {\bibfnamefont {J.-W.}\ \bibnamefont
  {Pan}}, \bibinfo {author} {\bibfnamefont {D.}~\bibnamefont {Bouwmeester}},
  \bibinfo {author} {\bibfnamefont {H.}~\bibnamefont {Weinfurter}}, \ and\
  \bibinfo {author} {\bibfnamefont {A.}~\bibnamefont {Zeilinger}},\ }\href@noop
  {} {\bibfield  {journal} {\bibinfo  {journal} {Physical Review Letters}\
  }\textbf {\bibinfo {volume} {80}},\ \bibinfo {pages} {3891} (\bibinfo {year}
  {1998})}\BibitemShut {NoStop}%
\bibitem [{\citenamefont {Shi}\ \emph {et~al.}(2000)\citenamefont {Shi},
  \citenamefont {Jiang},\ and\ \citenamefont {Guo}}]{Shi2000}%
  \BibitemOpen
  \bibfield  {author} {\bibinfo {author} {\bibfnamefont {B.-S.}\ \bibnamefont
  {Shi}}, \bibinfo {author} {\bibfnamefont {Y.-K.}\ \bibnamefont {Jiang}}, \
  and\ \bibinfo {author} {\bibfnamefont {G.-C.}\ \bibnamefont {Guo}},\
  }\href@noop {} {\bibfield  {journal} {\bibinfo  {journal} {Physical Review
  A}\ }\textbf {\bibinfo {volume} {62}},\ \bibinfo {pages} {054301} (\bibinfo
  {year} {2000})}\BibitemShut {NoStop}%
\bibitem [{\citenamefont {Jennewein}\ \emph {et~al.}(2001)\citenamefont
  {Jennewein}, \citenamefont {Weihs}, \citenamefont {Pan},\ and\ \citenamefont
  {Zeilinger}}]{Jennewein2001}%
  \BibitemOpen
  \bibfield  {author} {\bibinfo {author} {\bibfnamefont {T.}~\bibnamefont
  {Jennewein}}, \bibinfo {author} {\bibfnamefont {G.}~\bibnamefont {Weihs}},
  \bibinfo {author} {\bibfnamefont {J.-W.}\ \bibnamefont {Pan}}, \ and\
  \bibinfo {author} {\bibfnamefont {A.}~\bibnamefont {Zeilinger}},\ }\href@noop
  {} {\bibfield  {journal} {\bibinfo  {journal} {Physical Review Letters}\
  }\textbf {\bibinfo {volume} {88}},\ \bibinfo {pages} {017903} (\bibinfo
  {year} {2001})}\BibitemShut {NoStop}%
\bibitem [{\citenamefont {de~Riedmatten}\ \emph {et~al.}(2005)\citenamefont
  {de~Riedmatten}, \citenamefont {Marcikic}, \citenamefont {van Houwelingen},
  \citenamefont {Tittel}, \citenamefont {Zbinden},\ and\ \citenamefont
  {Gisin}}]{deRiedmatten2005}%
  \BibitemOpen
  \bibfield  {author} {\bibinfo {author} {\bibfnamefont {H.}~\bibnamefont
  {de~Riedmatten}}, \bibinfo {author} {\bibfnamefont {I.}~\bibnamefont
  {Marcikic}}, \bibinfo {author} {\bibfnamefont {J.~A.~W.}\ \bibnamefont {van
  Houwelingen}}, \bibinfo {author} {\bibfnamefont {W.}~\bibnamefont {Tittel}},
  \bibinfo {author} {\bibfnamefont {H.}~\bibnamefont {Zbinden}}, \ and\
  \bibinfo {author} {\bibfnamefont {N.}~\bibnamefont {Gisin}},\ }\href@noop {}
  {\bibfield  {journal} {\bibinfo  {journal} {Physical Review A}\ }\textbf
  {\bibinfo {volume} {71}},\ \bibinfo {pages} {050302} (\bibinfo {year}
  {2005})}\BibitemShut {NoStop}%
\bibitem [{\citenamefont {Kaltenbaek}\ \emph {et~al.}(2009)\citenamefont
  {Kaltenbaek}, \citenamefont {Prevedel}, \citenamefont {Aspelmeyer},\ and\
  \citenamefont {Zeilinger}}]{Kaltenbaek2009}%
  \BibitemOpen
  \bibfield  {author} {\bibinfo {author} {\bibfnamefont {R.}~\bibnamefont
  {Kaltenbaek}}, \bibinfo {author} {\bibfnamefont {R.}~\bibnamefont
  {Prevedel}}, \bibinfo {author} {\bibfnamefont {M.}~\bibnamefont
  {Aspelmeyer}}, \ and\ \bibinfo {author} {\bibfnamefont {A.}~\bibnamefont
  {Zeilinger}},\ }\href@noop {} {\bibfield  {journal} {\bibinfo  {journal}
  {Physical Review A}\ }\textbf {\bibinfo {volume} {79}},\ \bibinfo {pages}
  {040302} (\bibinfo {year} {2009})}\BibitemShut {NoStop}%
\bibitem [{\citenamefont {Ma}\ \emph {et~al.}(2012)\citenamefont {Ma},
  \citenamefont {Zotter}, \citenamefont {Kofler}, \citenamefont {Ursin},
  \citenamefont {Jennewein}, \citenamefont {Brukner},\ and\ \citenamefont
  {Zeilinger}}]{Ma2012}%
  \BibitemOpen
  \bibfield  {author} {\bibinfo {author} {\bibfnamefont {X.-S.}\ \bibnamefont
  {Ma}}, \bibinfo {author} {\bibfnamefont {S.}~\bibnamefont {Zotter}}, \bibinfo
  {author} {\bibfnamefont {J.}~\bibnamefont {Kofler}}, \bibinfo {author}
  {\bibfnamefont {R.}~\bibnamefont {Ursin}}, \bibinfo {author} {\bibfnamefont
  {T.}~\bibnamefont {Jennewein}}, \bibinfo {author} {\bibfnamefont {{\v
  C}.}~\bibnamefont {Brukner}}, \ and\ \bibinfo {author} {\bibfnamefont
  {A.}~\bibnamefont {Zeilinger}},\ }\href@noop {} {\bibfield  {journal}
  {\bibinfo  {journal} {Nature Physics}\ }\textbf {\bibinfo {volume} {8}},\
  \bibinfo {pages} {479} (\bibinfo {year} {2012})}\BibitemShut {NoStop}%
\bibitem [{\citenamefont {Peeters}\ \emph {et~al.}(2007)\citenamefont
  {Peeters}, \citenamefont {Verstegen},\ and\ \citenamefont {van
  Exter}}]{Peeters2007}%
  \BibitemOpen
  \bibfield  {author} {\bibinfo {author} {\bibfnamefont {W.~H.}\ \bibnamefont
  {Peeters}}, \bibinfo {author} {\bibfnamefont {E.~J.~K.}\ \bibnamefont
  {Verstegen}}, \ and\ \bibinfo {author} {\bibfnamefont {M.~P.}\ \bibnamefont
  {van Exter}},\ }\href@noop {} {\bibfield  {journal} {\bibinfo  {journal}
  {Physical Review A}\ }\textbf {\bibinfo {volume} {76}},\ \bibinfo {pages}
  {042302} (\bibinfo {year} {2007})}\BibitemShut {NoStop}%
\bibitem [{\citenamefont {Nagali}\ \emph {et~al.}(2009)\citenamefont {Nagali},
  \citenamefont {Sansoni}, \citenamefont {Sciarrino}, \citenamefont
  {De~Martini}, \citenamefont {Marrucci}, \citenamefont {Piccirillo},
  \citenamefont {Karimi},\ and\ \citenamefont {Santamato}}]{Nagali2009}%
  \BibitemOpen
  \bibfield  {author} {\bibinfo {author} {\bibfnamefont {E.}~\bibnamefont
  {Nagali}}, \bibinfo {author} {\bibfnamefont {L.}~\bibnamefont {Sansoni}},
  \bibinfo {author} {\bibfnamefont {F.}~\bibnamefont {Sciarrino}}, \bibinfo
  {author} {\bibfnamefont {F.}~\bibnamefont {De~Martini}}, \bibinfo {author}
  {\bibfnamefont {L.}~\bibnamefont {Marrucci}}, \bibinfo {author}
  {\bibfnamefont {B.}~\bibnamefont {Piccirillo}}, \bibinfo {author}
  {\bibfnamefont {E.}~\bibnamefont {Karimi}}, \ and\ \bibinfo {author}
  {\bibfnamefont {E.}~\bibnamefont {Santamato}},\ }\href@noop {} {\bibfield
  {journal} {\bibinfo  {journal} {Nature Photonics}\ }\textbf {\bibinfo
  {volume} {3}},\ \bibinfo {pages} {720} (\bibinfo {year} {2009})}\BibitemShut
  {NoStop}%
\bibitem [{\citenamefont {Di~Lorenzo~Pires}\ \emph {et~al.}(2010)\citenamefont
  {Di~Lorenzo~Pires}, \citenamefont {Florijn},\ and\ \citenamefont {van
  Exter}}]{DiLorenzoPires2010}%
  \BibitemOpen
  \bibfield  {author} {\bibinfo {author} {\bibfnamefont {H.}~\bibnamefont
  {Di~Lorenzo~Pires}}, \bibinfo {author} {\bibfnamefont {H.~C.~B.}\
  \bibnamefont {Florijn}}, \ and\ \bibinfo {author} {\bibfnamefont {M.~P.}\
  \bibnamefont {van Exter}},\ }\href@noop {} {\bibfield  {journal} {\bibinfo
  {journal} {Physical Review Letters}\ }\textbf {\bibinfo {volume} {104}},\
  \bibinfo {pages} {020505} (\bibinfo {year} {2010})}\BibitemShut {NoStop}%
\bibitem [{\citenamefont {Zhang}\ \emph {et~al.}(2016)\citenamefont {Zhang},
  \citenamefont {Roux}, \citenamefont {Konrad}, \citenamefont {Agnew},
  \citenamefont {Leach},\ and\ \citenamefont {Forbes}}]{Zhang2016}%
  \BibitemOpen
  \bibfield  {author} {\bibinfo {author} {\bibfnamefont {Y.}~\bibnamefont
  {Zhang}}, \bibinfo {author} {\bibfnamefont {F.~S.}\ \bibnamefont {Roux}},
  \bibinfo {author} {\bibfnamefont {T.}~\bibnamefont {Konrad}}, \bibinfo
  {author} {\bibfnamefont {M.}~\bibnamefont {Agnew}}, \bibinfo {author}
  {\bibfnamefont {J.}~\bibnamefont {Leach}}, \ and\ \bibinfo {author}
  {\bibfnamefont {A.}~\bibnamefont {Forbes}},\ }\href@noop {} {\bibfield
  {journal} {\bibinfo  {journal} {Science Advances}\ }\textbf {\bibinfo
  {volume} {2}},\ \bibinfo {pages} {e1501165} (\bibinfo {year}
  {2016})}\BibitemShut {NoStop}%
\bibitem [{\citenamefont {Leach}\ \emph {et~al.}(2010)\citenamefont {Leach},
  \citenamefont {Jack}, \citenamefont {Romero}, \citenamefont {Jha},
  \citenamefont {Yao}, \citenamefont {Franke-Arnold}, \citenamefont {Ireland},
  \citenamefont {Boyd}, \citenamefont {Barnett},\ and\ \citenamefont
  {Padgett}}]{Leach2010}%
  \BibitemOpen
  \bibfield  {author} {\bibinfo {author} {\bibfnamefont {J.}~\bibnamefont
  {Leach}}, \bibinfo {author} {\bibfnamefont {B.}~\bibnamefont {Jack}},
  \bibinfo {author} {\bibfnamefont {J.}~\bibnamefont {Romero}}, \bibinfo
  {author} {\bibfnamefont {A.~K.}\ \bibnamefont {Jha}}, \bibinfo {author}
  {\bibfnamefont {A.~M.}\ \bibnamefont {Yao}}, \bibinfo {author} {\bibfnamefont
  {S.}~\bibnamefont {Franke-Arnold}}, \bibinfo {author} {\bibfnamefont {D.~G.}\
  \bibnamefont {Ireland}}, \bibinfo {author} {\bibfnamefont {R.~W.}\
  \bibnamefont {Boyd}}, \bibinfo {author} {\bibfnamefont {S.~M.}\ \bibnamefont
  {Barnett}}, \ and\ \bibinfo {author} {\bibfnamefont {M.~J.}\ \bibnamefont
  {Padgett}},\ }\href@noop {} {\bibfield  {journal} {\bibinfo  {journal}
  {Science}\ }\textbf {\bibinfo {volume} {329}},\ \bibinfo {pages} {662}
  (\bibinfo {year} {2010})}\BibitemShut {NoStop}%
\bibitem [{\citenamefont {Agnew}\ \emph {et~al.}(2011)\citenamefont {Agnew},
  \citenamefont {Leach}, \citenamefont {McLaren}, \citenamefont {Roux},\ and\
  \citenamefont {Boyd}}]{Agnew2011}%
  \BibitemOpen
  \bibfield  {author} {\bibinfo {author} {\bibfnamefont {M.}~\bibnamefont
  {Agnew}}, \bibinfo {author} {\bibfnamefont {J.}~\bibnamefont {Leach}},
  \bibinfo {author} {\bibfnamefont {M.}~\bibnamefont {McLaren}}, \bibinfo
  {author} {\bibfnamefont {F.~S.}\ \bibnamefont {Roux}}, \ and\ \bibinfo
  {author} {\bibfnamefont {R.~W.}\ \bibnamefont {Boyd}},\ }\href@noop {}
  {\bibfield  {journal} {\bibinfo  {journal} {Physical Review A}\ }\textbf
  {\bibinfo {volume} {84}},\ \bibinfo {pages} {062101} (\bibinfo {year}
  {2011})}\BibitemShut {NoStop}%
\bibitem [{\citenamefont {Dada}\ \emph {et~al.}(2011)\citenamefont {Dada},
  \citenamefont {Leach}, \citenamefont {Buller}, \citenamefont {Padgett},\ and\
  \citenamefont {Andersson}}]{Dada2011}%
  \BibitemOpen
  \bibfield  {author} {\bibinfo {author} {\bibfnamefont {A.~C.}\ \bibnamefont
  {Dada}}, \bibinfo {author} {\bibfnamefont {J.}~\bibnamefont {Leach}},
  \bibinfo {author} {\bibfnamefont {G.~S.}\ \bibnamefont {Buller}}, \bibinfo
  {author} {\bibfnamefont {M.~J.}\ \bibnamefont {Padgett}}, \ and\ \bibinfo
  {author} {\bibfnamefont {E.}~\bibnamefont {Andersson}},\ }\href@noop {}
  {\bibfield  {journal} {\bibinfo  {journal} {Nature Physics}\ }\textbf
  {\bibinfo {volume} {7}},\ \bibinfo {pages} {677} (\bibinfo {year}
  {2011})}\BibitemShut {NoStop}%
\bibitem [{\citenamefont {Krenn}\ \emph {et~al.}(2014)\citenamefont {Krenn},
  \citenamefont {Huber}, \citenamefont {Fickler}, \citenamefont {Lapkiewicz},
  \citenamefont {Ramelow},\ and\ \citenamefont {Zeilinger}}]{Krenn2014}%
  \BibitemOpen
  \bibfield  {author} {\bibinfo {author} {\bibfnamefont {M.}~\bibnamefont
  {Krenn}}, \bibinfo {author} {\bibfnamefont {M.}~\bibnamefont {Huber}},
  \bibinfo {author} {\bibfnamefont {R.}~\bibnamefont {Fickler}}, \bibinfo
  {author} {\bibfnamefont {R.}~\bibnamefont {Lapkiewicz}}, \bibinfo {author}
  {\bibfnamefont {S.}~\bibnamefont {Ramelow}}, \ and\ \bibinfo {author}
  {\bibfnamefont {A.}~\bibnamefont {Zeilinger}},\ }\href@noop {} {\bibfield
  {journal} {\bibinfo  {journal} {Proceedings of the National Academy of
  Sciences}\ }\textbf {\bibinfo {volume} {111}},\ \bibinfo {pages} {6243}
  (\bibinfo {year} {2014})}\BibitemShut {NoStop}%
\bibitem [{\citenamefont {Bolduc}\ \emph {et~al.}(2016)\citenamefont {Bolduc},
  \citenamefont {Gariepy},\ and\ \citenamefont {Leach}}]{Bolduc2016}%
  \BibitemOpen
  \bibfield  {author} {\bibinfo {author} {\bibfnamefont {E.}~\bibnamefont
  {Bolduc}}, \bibinfo {author} {\bibfnamefont {G.}~\bibnamefont {Gariepy}}, \
  and\ \bibinfo {author} {\bibfnamefont {J.}~\bibnamefont {Leach}},\
  }\href@noop {} {\bibfield  {journal} {\bibinfo  {journal} {Nature
  Communications}\ }\textbf {\bibinfo {volume} {7}},\ \bibinfo {pages} {10439}
  (\bibinfo {year} {2016})}\BibitemShut {NoStop}%
\bibitem [{\citenamefont {Mair}\ \emph {et~al.}(2001)\citenamefont {Mair},
  \citenamefont {Vaziri}, \citenamefont {Weihs},\ and\ \citenamefont
  {Zeilinger}}]{Mair2001}%
  \BibitemOpen
  \bibfield  {author} {\bibinfo {author} {\bibfnamefont {A.}~\bibnamefont
  {Mair}}, \bibinfo {author} {\bibfnamefont {A.}~\bibnamefont {Vaziri}},
  \bibinfo {author} {\bibfnamefont {G.}~\bibnamefont {Weihs}}, \ and\ \bibinfo
  {author} {\bibfnamefont {A.}~\bibnamefont {Zeilinger}},\ }\href@noop {}
  {\bibfield  {journal} {\bibinfo  {journal} {Nature}\ }\textbf {\bibinfo
  {volume} {412}},\ \bibinfo {pages} {313} (\bibinfo {year}
  {2001})}\BibitemShut {NoStop}%
\bibitem [{\citenamefont {Walborn}\ \emph {et~al.}(2004)\citenamefont
  {Walborn}, \citenamefont {de~Oliveira}, \citenamefont {Thebaldi},\ and\
  \citenamefont {Monken}}]{Walborn2004}%
  \BibitemOpen
  \bibfield  {author} {\bibinfo {author} {\bibfnamefont {S.~P.}\ \bibnamefont
  {Walborn}}, \bibinfo {author} {\bibfnamefont {A.~N.}\ \bibnamefont
  {de~Oliveira}}, \bibinfo {author} {\bibfnamefont {R.~S.}\ \bibnamefont
  {Thebaldi}}, \ and\ \bibinfo {author} {\bibfnamefont {C.~H.}\ \bibnamefont
  {Monken}},\ }\href@noop {} {\bibfield  {journal} {\bibinfo  {journal}
  {Physical Review A}\ }\textbf {\bibinfo {volume} {69}},\ \bibinfo {pages}
  {023811} (\bibinfo {year} {2004})}\BibitemShut {NoStop}%
\bibitem [{\citenamefont {Mafu}\ \emph {et~al.}(2013)\citenamefont {Mafu},
  \citenamefont {Dudley}, \citenamefont {Goyal}, \citenamefont {Giovannini},
  \citenamefont {McLaren}, \citenamefont {Padgett}, \citenamefont {Konrad},
  \citenamefont {Petruccione}, \citenamefont {L\"utkenhaus},\ and\
  \citenamefont {Forbes}}]{Mafu2013}%
  \BibitemOpen
  \bibfield  {author} {\bibinfo {author} {\bibfnamefont {M.}~\bibnamefont
  {Mafu}}, \bibinfo {author} {\bibfnamefont {A.}~\bibnamefont {Dudley}},
  \bibinfo {author} {\bibfnamefont {S.}~\bibnamefont {Goyal}}, \bibinfo
  {author} {\bibfnamefont {D.}~\bibnamefont {Giovannini}}, \bibinfo {author}
  {\bibfnamefont {M.}~\bibnamefont {McLaren}}, \bibinfo {author} {\bibfnamefont
  {M.~J.}\ \bibnamefont {Padgett}}, \bibinfo {author} {\bibfnamefont
  {T.}~\bibnamefont {Konrad}}, \bibinfo {author} {\bibfnamefont
  {F.}~\bibnamefont {Petruccione}}, \bibinfo {author} {\bibfnamefont
  {N.}~\bibnamefont {L\"utkenhaus}}, \ and\ \bibinfo {author} {\bibfnamefont
  {A.}~\bibnamefont {Forbes}},\ }\href {\doibase 10.1103/PhysRevA.88.032305}
  {\bibfield  {journal} {\bibinfo  {journal} {Phys. Rev. A}\ }\textbf {\bibinfo
  {volume} {88}},\ \bibinfo {pages} {032305} (\bibinfo {year}
  {2013})}\BibitemShut {NoStop}%
\bibitem [{\citenamefont {Mirhosseini}\ \emph {et~al.}(2015)\citenamefont
  {Mirhosseini}, \citenamefont {Maga{\~n}a-Loaiza}, \citenamefont {O'Sullivan},
  \citenamefont {Rodenburg}, \citenamefont {Malik}, \citenamefont {Lavery},
  \citenamefont {Padgett}, \citenamefont {Gauthier},\ and\ \citenamefont
  {Boyd}}]{Mirhosseini2015}%
  \BibitemOpen
  \bibfield  {author} {\bibinfo {author} {\bibfnamefont {M.}~\bibnamefont
  {Mirhosseini}}, \bibinfo {author} {\bibfnamefont {O.~S.}\ \bibnamefont
  {Maga{\~n}a-Loaiza}}, \bibinfo {author} {\bibfnamefont {M.~N.}\ \bibnamefont
  {O'Sullivan}}, \bibinfo {author} {\bibfnamefont {B.}~\bibnamefont
  {Rodenburg}}, \bibinfo {author} {\bibfnamefont {M.}~\bibnamefont {Malik}},
  \bibinfo {author} {\bibfnamefont {M.~P.~J.}\ \bibnamefont {Lavery}}, \bibinfo
  {author} {\bibfnamefont {M.~J.}\ \bibnamefont {Padgett}}, \bibinfo {author}
  {\bibfnamefont {D.~J.}\ \bibnamefont {Gauthier}}, \ and\ \bibinfo {author}
  {\bibfnamefont {R.~W.}\ \bibnamefont {Boyd}},\ }\href
  {http://stacks.iop.org/1367-2630/17/i=3/a=033033} {\bibfield  {journal}
  {\bibinfo  {journal} {New Journal of Physics}\ }\textbf {\bibinfo {volume}
  {17}},\ \bibinfo {pages} {033033} (\bibinfo {year} {2015})}\BibitemShut
  {NoStop}%
\bibitem [{\citenamefont {Marcikic}\ \emph {et~al.}(2002)\citenamefont
  {Marcikic}, \citenamefont {de~Riedmatten}, \citenamefont {Tittel},
  \citenamefont {Scarani}, \citenamefont {Zbinden},\ and\ \citenamefont
  {Gisin}}]{Marcikic2002}%
  \BibitemOpen
  \bibfield  {author} {\bibinfo {author} {\bibfnamefont {I.}~\bibnamefont
  {Marcikic}}, \bibinfo {author} {\bibfnamefont {H.}~\bibnamefont
  {de~Riedmatten}}, \bibinfo {author} {\bibfnamefont {W.}~\bibnamefont
  {Tittel}}, \bibinfo {author} {\bibfnamefont {V.}~\bibnamefont {Scarani}},
  \bibinfo {author} {\bibfnamefont {H.}~\bibnamefont {Zbinden}}, \ and\
  \bibinfo {author} {\bibfnamefont {N.}~\bibnamefont {Gisin}},\ }\href@noop {}
  {\bibfield  {journal} {\bibinfo  {journal} {Physical Review A}\ }\textbf
  {\bibinfo {volume} {66}},\ \bibinfo {pages} {062308} (\bibinfo {year}
  {2002})}\BibitemShut {NoStop}%
\bibitem [{\citenamefont {Matthews}\ \emph {et~al.}(2009)\citenamefont
  {Matthews}, \citenamefont {Politi}, \citenamefont {Stefanov},\ and\
  \citenamefont {O'Brien}}]{Matthews2009}%
  \BibitemOpen
  \bibfield  {author} {\bibinfo {author} {\bibfnamefont {J.~C.~F.}\
  \bibnamefont {Matthews}}, \bibinfo {author} {\bibfnamefont {A.}~\bibnamefont
  {Politi}}, \bibinfo {author} {\bibfnamefont {A.}~\bibnamefont {Stefanov}}, \
  and\ \bibinfo {author} {\bibfnamefont {J.~L.}\ \bibnamefont {O'Brien}},\
  }\href@noop {} {\bibfield  {journal} {\bibinfo  {journal} {Nature Photonics}\
  }\textbf {\bibinfo {volume} {3}},\ \bibinfo {pages} {346} (\bibinfo {year}
  {2009})}\BibitemShut {NoStop}%
\bibitem [{\citenamefont {Barreiro}\ \emph {et~al.}(2008)\citenamefont
  {Barreiro}, \citenamefont {Wei},\ and\ \citenamefont {Kwiat}}]{Barreiro2008}%
  \BibitemOpen
  \bibfield  {author} {\bibinfo {author} {\bibfnamefont {J.~T.}\ \bibnamefont
  {Barreiro}}, \bibinfo {author} {\bibfnamefont {T.-C.}\ \bibnamefont {Wei}}, \
  and\ \bibinfo {author} {\bibfnamefont {P.~G.}\ \bibnamefont {Kwiat}},\
  }\href@noop {} {\bibfield  {journal} {\bibinfo  {journal} {Nature Physics}\
  }\textbf {\bibinfo {volume} {4}},\ \bibinfo {pages} {282} (\bibinfo {year}
  {2008})}\BibitemShut {NoStop}%
\bibitem [{\citenamefont {Karimi}\ \emph {et~al.}(2010)\citenamefont {Karimi},
  \citenamefont {Leach}, \citenamefont {Slussarenko}, \citenamefont
  {Piccirillo}, \citenamefont {Marrucci}, \citenamefont {Chen}, \citenamefont
  {She}, \citenamefont {Franke-Arnold}, \citenamefont {Padgett},\ and\
  \citenamefont {Santamato}}]{Karimi2010}%
  \BibitemOpen
  \bibfield  {author} {\bibinfo {author} {\bibfnamefont {E.}~\bibnamefont
  {Karimi}}, \bibinfo {author} {\bibfnamefont {J.}~\bibnamefont {Leach}},
  \bibinfo {author} {\bibfnamefont {S.}~\bibnamefont {Slussarenko}}, \bibinfo
  {author} {\bibfnamefont {B.}~\bibnamefont {Piccirillo}}, \bibinfo {author}
  {\bibfnamefont {L.}~\bibnamefont {Marrucci}}, \bibinfo {author}
  {\bibfnamefont {L.}~\bibnamefont {Chen}}, \bibinfo {author} {\bibfnamefont
  {W.}~\bibnamefont {She}}, \bibinfo {author} {\bibfnamefont {S.}~\bibnamefont
  {Franke-Arnold}}, \bibinfo {author} {\bibfnamefont {M.~J.}\ \bibnamefont
  {Padgett}}, \ and\ \bibinfo {author} {\bibfnamefont {E.}~\bibnamefont
  {Santamato}},\ }\href@noop {} {\bibfield  {journal} {\bibinfo  {journal}
  {Physical Review A}\ }\textbf {\bibinfo {volume} {82}},\ \bibinfo {pages}
  {022115} (\bibinfo {year} {2010})}\BibitemShut {NoStop}%
\bibitem [{\citenamefont {Wang}\ \emph {et~al.}(2015)\citenamefont {Wang},
  \citenamefont {Cai}, \citenamefont {Su}, \citenamefont {Chen}, \citenamefont
  {Wu}, \citenamefont {Li}, \citenamefont {Liu}, \citenamefont {Lu},\ and\
  \citenamefont {Pan}}]{Wang2015}%
  \BibitemOpen
  \bibfield  {author} {\bibinfo {author} {\bibfnamefont {X.-L.}\ \bibnamefont
  {Wang}}, \bibinfo {author} {\bibfnamefont {X.-D.}\ \bibnamefont {Cai}},
  \bibinfo {author} {\bibfnamefont {Z.-E.}\ \bibnamefont {Su}}, \bibinfo
  {author} {\bibfnamefont {M.-C.}\ \bibnamefont {Chen}}, \bibinfo {author}
  {\bibfnamefont {D.}~\bibnamefont {Wu}}, \bibinfo {author} {\bibfnamefont
  {L.}~\bibnamefont {Li}}, \bibinfo {author} {\bibfnamefont {N.-L.}\
  \bibnamefont {Liu}}, \bibinfo {author} {\bibfnamefont {C.-Y.}\ \bibnamefont
  {Lu}}, \ and\ \bibinfo {author} {\bibfnamefont {J.-W.}\ \bibnamefont {Pan}},\
  }\href@noop {} {\bibfield  {journal} {\bibinfo  {journal} {Nature}\ }\textbf
  {\bibinfo {volume} {518}},\ \bibinfo {pages} {516} (\bibinfo {year}
  {2015})}\BibitemShut {NoStop}%
\bibitem [{\citenamefont {Graham}\ \emph {et~al.}(2015)\citenamefont {Graham},
  \citenamefont {Bernstein}, \citenamefont {Wei}, \citenamefont {Junge},\ and\
  \citenamefont {Kwiat}}]{Graham2015}%
  \BibitemOpen
  \bibfield  {author} {\bibinfo {author} {\bibfnamefont {T.~M.}\ \bibnamefont
  {Graham}}, \bibinfo {author} {\bibfnamefont {H.~J.}\ \bibnamefont
  {Bernstein}}, \bibinfo {author} {\bibfnamefont {T.-C.}\ \bibnamefont {Wei}},
  \bibinfo {author} {\bibfnamefont {M.}~\bibnamefont {Junge}}, \ and\ \bibinfo
  {author} {\bibfnamefont {P.~G.}\ \bibnamefont {Kwiat}},\ }\href@noop {}
  {\bibfield  {journal} {\bibinfo  {journal} {Nature Communications}\ }\textbf
  {\bibinfo {volume} {6}},\ \bibinfo {pages} {7185} (\bibinfo {year}
  {2015})}\BibitemShut {NoStop}%
\bibitem [{\citenamefont {Hiesmayr}\ \emph {et~al.}(2016)\citenamefont
  {Hiesmayr}, \citenamefont {de~Dood},\ and\ \citenamefont
  {L{\"o}ffler}}]{Hiesmayr2016}%
  \BibitemOpen
  \bibfield  {author} {\bibinfo {author} {\bibfnamefont {B.~C.}\ \bibnamefont
  {Hiesmayr}}, \bibinfo {author} {\bibfnamefont {M.~J.~A.}\ \bibnamefont
  {de~Dood}}, \ and\ \bibinfo {author} {\bibfnamefont {W.}~\bibnamefont
  {L{\"o}ffler}},\ }\href@noop {} {\bibfield  {journal} {\bibinfo  {journal}
  {Physical Review Letters}\ }\textbf {\bibinfo {volume} {116}},\ \bibinfo
  {pages} {073601} (\bibinfo {year} {2016})}\BibitemShut {NoStop}%
\bibitem [{\citenamefont {Malik}\ \emph {et~al.}(2016)\citenamefont {Malik},
  \citenamefont {Erhard}, \citenamefont {Huber}, \citenamefont {Krenn},
  \citenamefont {Fickler},\ and\ \citenamefont {Zeilinger}}]{Malik2016}%
  \BibitemOpen
  \bibfield  {author} {\bibinfo {author} {\bibfnamefont {M.}~\bibnamefont
  {Malik}}, \bibinfo {author} {\bibfnamefont {M.}~\bibnamefont {Erhard}},
  \bibinfo {author} {\bibfnamefont {M.}~\bibnamefont {Huber}}, \bibinfo
  {author} {\bibfnamefont {M.}~\bibnamefont {Krenn}}, \bibinfo {author}
  {\bibfnamefont {R.}~\bibnamefont {Fickler}}, \ and\ \bibinfo {author}
  {\bibfnamefont {A.}~\bibnamefont {Zeilinger}},\ }\href@noop {} {\bibfield
  {journal} {\bibinfo  {journal} {Nature Photonics}\ }\textbf {\bibinfo
  {volume} {10}},\ \bibinfo {pages} {248} (\bibinfo {year} {2016})}\BibitemShut
  {NoStop}%
\bibitem [{\citenamefont {Goyal}\ and\ \citenamefont
  {Konrad}(2013)}]{Goyal2013}%
  \BibitemOpen
  \bibfield  {author} {\bibinfo {author} {\bibfnamefont {S.~K.}\ \bibnamefont
  {Goyal}}\ and\ \bibinfo {author} {\bibfnamefont {T.}~\bibnamefont {Konrad}},\
  }\href@noop {} {\bibfield  {journal} {\bibinfo  {journal} {Scientific
  Reports}\ }\textbf {\bibinfo {volume} {3}},\ \bibinfo {pages} {3548}
  (\bibinfo {year} {2013})}\BibitemShut {NoStop}%
\bibitem [{\citenamefont {Goyal}\ \emph {et~al.}(2014)\citenamefont {Goyal},
  \citenamefont {Boukama-Dzoussi}, \citenamefont {Ghosh}, \citenamefont
  {Roux},\ and\ \citenamefont {Konrad}}]{Goyal2014}%
  \BibitemOpen
  \bibfield  {author} {\bibinfo {author} {\bibfnamefont {S.~K.}\ \bibnamefont
  {Goyal}}, \bibinfo {author} {\bibfnamefont {P.~E.}\ \bibnamefont
  {Boukama-Dzoussi}}, \bibinfo {author} {\bibfnamefont {S.}~\bibnamefont
  {Ghosh}}, \bibinfo {author} {\bibfnamefont {F.~S.}\ \bibnamefont {Roux}}, \
  and\ \bibinfo {author} {\bibfnamefont {T.}~\bibnamefont {Konrad}},\
  }\href@noop {} {\bibfield  {journal} {\bibinfo  {journal} {Scientific
  Reports}\ }\textbf {\bibinfo {volume} {4}},\ \bibinfo {pages} {4543}
  (\bibinfo {year} {2014})}\BibitemShut {NoStop}%
\bibitem [{\citenamefont {Salakhutdinov}\ \emph {et~al.}(2012)\citenamefont
  {Salakhutdinov}, \citenamefont {Eliel},\ and\ \citenamefont
  {L{\"o}ffler}}]{Salakhutdinov2012}%
  \BibitemOpen
  \bibfield  {author} {\bibinfo {author} {\bibfnamefont {V.~D.}\ \bibnamefont
  {Salakhutdinov}}, \bibinfo {author} {\bibfnamefont {E.~R.}\ \bibnamefont
  {Eliel}}, \ and\ \bibinfo {author} {\bibfnamefont {W.}~\bibnamefont
  {L{\"o}ffler}},\ }\href@noop {} {\bibfield  {journal} {\bibinfo  {journal}
  {Physical Review Letters}\ }\textbf {\bibinfo {volume} {108}},\ \bibinfo
  {pages} {173604} (\bibinfo {year} {2012})}\BibitemShut {NoStop}%
\bibitem [{\citenamefont {Fickler}\ \emph {et~al.}(2012)\citenamefont
  {Fickler}, \citenamefont {Lapkiewicz}, \citenamefont {Plick}, \citenamefont
  {Krenn}, \citenamefont {Schaeff}, \citenamefont {Ramelow},\ and\
  \citenamefont {Zeilinger}}]{Fickler2012}%
  \BibitemOpen
  \bibfield  {author} {\bibinfo {author} {\bibfnamefont {R.}~\bibnamefont
  {Fickler}}, \bibinfo {author} {\bibfnamefont {R.}~\bibnamefont {Lapkiewicz}},
  \bibinfo {author} {\bibfnamefont {W.~N.}\ \bibnamefont {Plick}}, \bibinfo
  {author} {\bibfnamefont {M.}~\bibnamefont {Krenn}}, \bibinfo {author}
  {\bibfnamefont {C.}~\bibnamefont {Schaeff}}, \bibinfo {author} {\bibfnamefont
  {S.}~\bibnamefont {Ramelow}}, \ and\ \bibinfo {author} {\bibfnamefont
  {A.}~\bibnamefont {Zeilinger}},\ }\href@noop {} {\bibfield  {journal}
  {\bibinfo  {journal} {Review of Scientific Instruments}\ }\textbf {\bibinfo
  {volume} {338}},\ \bibinfo {pages} {640} (\bibinfo {year}
  {2012})}\BibitemShut {NoStop}%
\bibitem [{\citenamefont {Guerreiro}\ \emph {et~al.}(2013)\citenamefont
  {Guerreiro}, \citenamefont {Pomarico}, \citenamefont {Sanguinetti},
  \citenamefont {Sangouard}, \citenamefont {Pelc}, \citenamefont {Langrock},
  \citenamefont {Fejer}, \citenamefont {Zbinden}, \citenamefont {Thew},\ and\
  \citenamefont {Gisin}}]{Guerreiro2013}%
  \BibitemOpen
  \bibfield  {author} {\bibinfo {author} {\bibfnamefont {T.}~\bibnamefont
  {Guerreiro}}, \bibinfo {author} {\bibfnamefont {E.}~\bibnamefont {Pomarico}},
  \bibinfo {author} {\bibfnamefont {B.}~\bibnamefont {Sanguinetti}}, \bibinfo
  {author} {\bibfnamefont {N.}~\bibnamefont {Sangouard}}, \bibinfo {author}
  {\bibfnamefont {J.~S.}\ \bibnamefont {Pelc}}, \bibinfo {author}
  {\bibfnamefont {C.}~\bibnamefont {Langrock}}, \bibinfo {author}
  {\bibfnamefont {M.~M.}\ \bibnamefont {Fejer}}, \bibinfo {author}
  {\bibfnamefont {H.}~\bibnamefont {Zbinden}}, \bibinfo {author} {\bibfnamefont
  {R.~T.}\ \bibnamefont {Thew}}, \ and\ \bibinfo {author} {\bibfnamefont
  {N.}~\bibnamefont {Gisin}},\ }\href@noop {} {\bibfield  {journal} {\bibinfo
  {journal} {Nature Communications}\ }\textbf {\bibinfo {volume} {4}},\
  \bibinfo {pages} {2324} (\bibinfo {year} {2013})}\BibitemShut {NoStop}%
\end{thebibliography}
\end{document}